\documentclass{article}

\usepackage{arxiv}
\usepackage{cite}
\usepackage{amsmath,amssymb,amsfonts}
\usepackage{algorithmic}
\usepackage{graphicx}
\usepackage{textcomp}
\usepackage{xcolor}
\usepackage[utf8]{inputenc}
\usepackage{caption}
\usepackage{subcaption}
\usepackage{hyperref}
\usepackage{url}
\usepackage{adjustbox}
\usepackage{multirow}
\usepackage{multicol}
\usepackage{array}
\usepackage{color, colortbl}
\usepackage{makecell}
\usepackage{hhline}
\usepackage{flushend}
\usepackage{rotating}
\usepackage{booktabs} 
\definecolor{Gray}{gray}{0.9}
\definecolor{LightCyan}{rgb}{0.88,1,1}

\usepackage{makecell}

\title{On Technical Trading and Social Media Indicators in Cryptocurrency Price Classification Through Deep Learning}

\author{Marco Ortu \\
            University of Cagliari\\
            marco.ortu@unica.it
        \and Nicola Uras \\
            University of Cagliari\\
            nicola.uras@unica.it
        \and Claudio Conversano \\
            University of Cagliari\\
            conversano@unica.it \vspace{0.4cm}
        \and Giuseppe Destefanis \\
            Brunel University\\
            giuseppe.destefanis@brunel.ac.uk
        \and Silvia Bartolucci\\
            University College London\\
             s.bartolucci@ucl.ac.uk
}

\date{January 2021}

\begin{document}

\maketitle

\begin{abstract}
Predicting prices of cryptocurrencies is a notoriously hard task due to the presence of high volatility and new mechanisms characterising the crypto markets. 
In this work we focus on the two major cryptocurrencies for market capitalization at the time of the study, Ethereum and Bitcoin, for the period 2017-2020.
We present a comprehensive analysis of the predictability of price movements comparing four different deep learning algorithms (\textit{Multi Layers Perceptron (MLP)},  \textit{Convolutional Neural Network (CNN)}, \textit{Long Short Term Memory (LSTM) neural network} and \textit{Attention Long Short Term Memory (ALSTM)}) and using three classes of features. In particular, we consider a combination of technical (e.g. open and close prices), trading (e.g. moving averages) and social (e.g. users' sentiment) indicators used as input to our classification algorithm.
We compare a \textit{restricted model} composed of technical indicators only, and an \textit{unrestricted model} including technical, trading and social media indicators.
The results show that the unrestricted model outperforms the restricted one, i.e. including trading and social media indicators, along with the classic technical variables, leads to a significant improvement in the prediction accuracy consistently across all algorithms. 

\end{abstract}


\section{Introduction}
\label{sec:introduction}

During the last decade, the global markets have witnessed the rise and exponential growth of cryptocurrencies traded and exchanged with a daily market capitalization of hundreds of billions of USD Dollars globally (reaching $\approx$ 1 trillion as of January 2021). 

Recent surveys\footnote{See \href{https://www.fidelitydigitalassets.com/bin-public/060_www_fidelity_com/documents/FDAS/institutional-investors-digital-asset-survey.pdf}{Fidelity Report.}} report a spike in demand and interest for the new crypto assets from institutional investors, attracted by the novel features and the potential rise in value in the current financial turmoil, despite the risk associated with price volatility and market manipulation.

Boom and bust cycles often induced by network effects and wider market's adoption, make prices hard to predict with high accuracy. There is a large body of literature concerning this issue and proposing a number of quantitative approaches for cryptocurrency prices prediction \cite{karim2019multivariate, NNprediction,returnsprediction,katsiampa2017volatility,lahmiri2018long}. The rapid fluctuations in volatility, autocorrelations and multi-scaling effects in cryptocurrencies have also been extensively studied \cite{matta2015bitcoin}, also with respect to their effect on Initial Coin Offering (ICO) \cite{hartmann2018evaluation, hartmann2019alternative}.

An important consideration that has gradually emerged from the literature is the relevance of the ``social aspect" of crypto trading. The code underlying blockchain platforms is developed in an open-source fashion on Github, recent additions to the crypto ecosystem are discussed on Reddit or on specialised channels in Telegram, and Twitter offers a platform where often heated debates on the latest developments take place. More precisely, it has been shown that sentiment index can be used to predict  bubbles in prices \cite{chensentiment} and that the sentiment extracted from topic discussions on Reddit correlates with prices \cite{phillips2018mutual}. 

Open-source development also plays an important role in shaping the success and value of cryptocurrencies \cite{ortu2015mining,ortu2019comparing, marchesi2020design}. In particular, a previous work by Bartolucci et al. \cite{bartolucci2020butterfly} -- which this work is an extension of -- showed the existence of a Granger causality between the sentiment and emotions time series extracted from developers' comments on Github and returns of cryptocurrencies. For the two major cryptocurrencies -- Bitcoin and Ethereum -- it has been also shown how including the developers' emotions time series in prediction algorithms could substantially improve the accuracy. 

In this paper, we further extend previous investigations on price predictability using a deep learning approach and focusing on the two major cryptocurrencies by market capitalization, Bitcoin and Ethereum.

We predict price movements by mapping the punctual price forecasting into a classification problem: our target is a binary variable with two unique classes, upward and downward movements, which indicate prices rising or falling. In the following we will compare the performances and outcome of four deep learning algorithms: the Multi-Layer Perceptron (MLP), the Multivariate Attention Long Short Term Memory Fully Convolutional Network (MALSTM-FCN), the Convolutional Neural Network (CNN) and the Long Short Term Memory neural network (LSTM). 

We will use as input the following classes of (financial and social) indicators: (i) technical indicators, such as open and close price or volume traded, (ii) trading indicators, such as the momentum and moving averages calculated on the price, (iii) social media indicators, i.e. sentiment and emotions extracted from Github and Reddit comments.

For each deep learning algorithm we consider a \textit{restricted} and \textit{unrestricted} data model at a hourly and daily frequency. The \textit{restricted model} consists of data concerning technical variables for Bitcoin and Ethereum. In the \textit{unrestricted model} we include, instead, the technical variables, trading and social media indicators from Github and Reddit. 
 
Consistently across all four deep learning algorithms, we are able to show that that the unrestricted model outperforms the restricted model. At hourly data frequency, the inclusion of  trading and social media indicators alongside the classic technical indicators improves the accuracy on Bitcoin and Ethereum price prediction, increasing from a range of 51-55\% for the restricted model to 67-84\% for the unrestricted one. For the daily frequency resolution, in the case of Ethereum the most accurate classification is achieved using the restricted model. For Bitcoin, instead, the highest performance is achieved for the unrestricted model including only social media indicators.

In the following sections we will discuss in details the algorithms implemented and the Bootstrap validation technique used to estimate the performance of the models.

The paper is organised as follows.
In Section \ref{sec:dateset} we describe in detail the data and indicators used. In Section \ref{sec:methodology}, we discuss the methodology of the experiments conducted. In Section \ref{sec:results} we present the results and their implications and in Section \ref{sec:threats} we discuss the limitations of this study. Finally, in Section \ref{sec:conclusions} we summarise our findings and outline future directions.

\section{Dataset: Technical and Social Media Indicators}
\label{sec:dateset}

This section discusses the dataset and the three categories of indicators used for the experiments.

\subsection{Technical Indicators}
\label{subsec:technical_trading_indicators}

We conducted our analysis on Bitcoin and Ethereum price time series with an hourly and daily frequency resolution.
We considered all the available technical variables, extracted from the \textit{Crypto Data Download} web services\footnote{https://www.cryptodatadownload.com/data/bitfinex/}, in particular the data from Bitfinex.com exchange\footnote{https://www.bitfinex.com/} service.
We considered the last 4-year period, spanning from $2017/01/01$ to $2021/01/01$, for a total of $35638$ hourly observations.

In our analysis, we separate the technical indicators into two main categories: pure technical and trading indicators. Technical indicators refer to ``direct" market data such as opening and closing prices. Trading indicators refer to derived indicators such as the moving averages.

The technical indicators are listed below.

\begin{itemize}
    \item \textit{Close}: the last price at which the cryptocurrency traded during the trading period.
    \item \textit{Open}: the price at which the cryptocurrency first trades upon the opening of a trading period.
    \item \textit{Low}: the lowest price at which the cryptocurrency trades over the course of a trading period.
    \item \textit{High}: the highest price at which the cryptocurrency traded during the course of the trading period.
    \item \textit{Volume}: the number of cryptocurrency trades completed.
\end{itemize}

\begin{figure}[htbp]
\centering
\includegraphics[width=\textwidth]{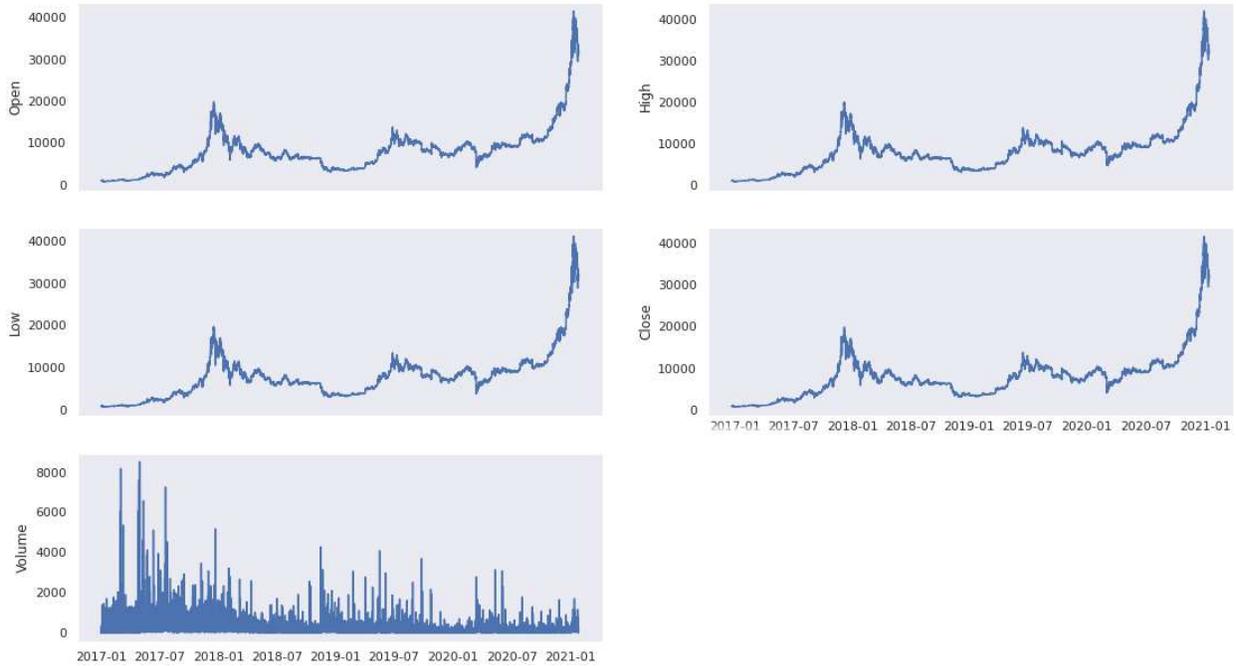}
\caption{Plot of the time series of Bitcoin's technical indicators.}
\label{figure:bitcoin_technical_metrics}
\end{figure}

\begin{figure}[htbp]
\centering
\includegraphics[width=\textwidth]{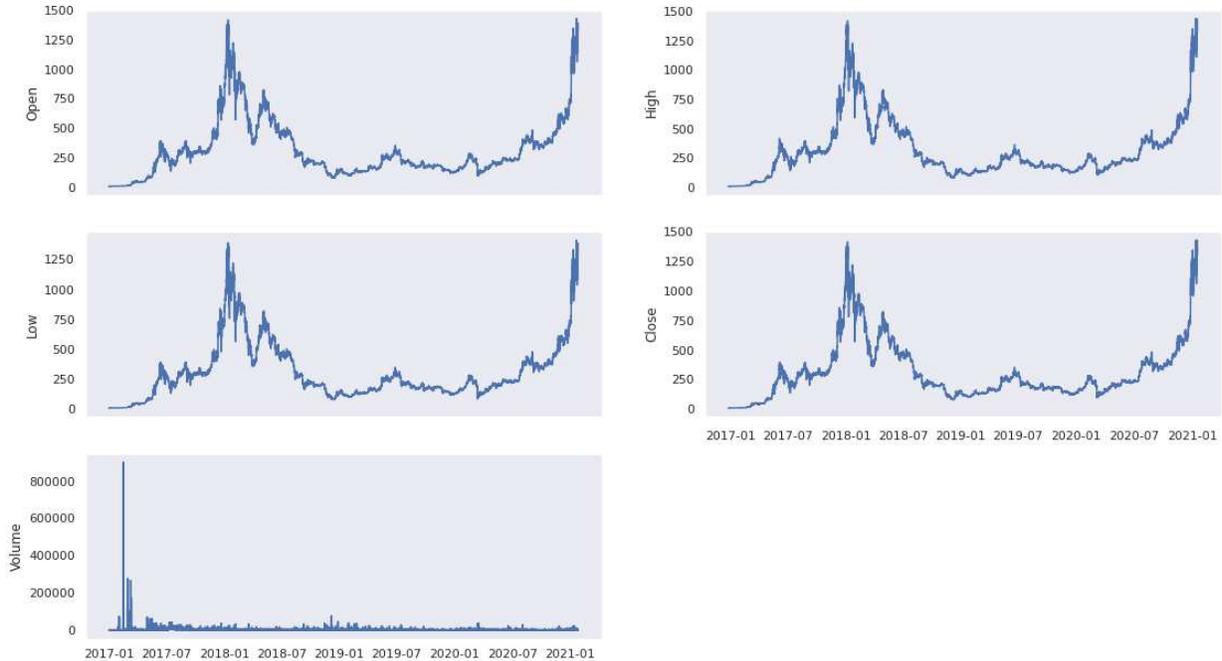}
\caption{Plot of the time series of Ethereum's technical indicators.}
\label{figure:ethereum_technical_metrics}
\end{figure}

Tables \ref{Table:bitcoin_technical_metrics} and \ref{Table:ethereum_technical_metrics} show the summary statistics for the technical indicators. In Figures \ref{figure:bitcoin_technical_metrics} and \ref{figure:ethereum_technical_metrics} we also show the plot of the historical time series for the technical indicators.

\begin{table}[htbp]
\centering
\begin{tabular}{lrrrrr}
\toprule
{} &          \textbf{High} &          \textbf{Open} &           \textbf{Low} &        \textbf{Volume} &         \textbf{Close}\\
\midrule
\textbf{mean}  &   7972.769 &   7928.018 &   7879.276 &    176.319 &   7928.894  \\
\textbf{std}   &   5519.337 &   5471.592 &   5416.295 &    306.62 &   5472.983  \\
\textbf{min}   &    769.1 &    760.38 &    752 &      0 &    760.38\\
\textbf{25\%}   &   4161.6875&   4137.995&   4113.822&     30.592 &   4138.475\\
\textbf{50\%}   &   7459.995 &   7428.09 &   7390.47 &     80.699 &   7428.41\\
\textbf{75\%}   &   9790.952&   9751.84 &   9701.427 &    199.322 &   9752.37\\
\textbf{max}   &  41999.99 &  41526.95 &  41000.24 &   8526.751 &  41526.95 \\
\bottomrule
\end{tabular}
\caption{Summary statistics for the time series of Bitcoin's technical indicators.}
\label{Table:bitcoin_technical_metrics}
\end{table}

\begin{table}[htbp]
\centering
\begin{tabular}{lrrrrr}
\toprule
{} &          \textbf{High} &          \textbf{Open} &           \textbf{Low} &        \textbf{Volume} &         \textbf{Close}\\
\midrule
\textbf{mean}  &    313.202 &    310.856 &    308.253 &    1658.835 &    310.896\\
\textbf{std}   &    248.731 &    246.069 &    242.971 &    6903.628 &    246.135 \\
\textbf{min}   &      8.17 &      8.15 &      8.15 &       0&      8.15\\
\textbf{25\%}   &    161.182 &    160.202 &    159.06 &     192.327 &    160.21 \\
\textbf{50\% }  &    232.79 &    231.34 &    229.765 &     569.79 &    231.365 \\
\textbf{75\%}   &    390.0575 &    388.0075 &    385.73 &    1632.640 &    388.025 \\
\textbf{max}   &   1440.54 &   1430.94 &   1411 &  903102.685 &   1431.4\\
\bottomrule
\end{tabular}
\caption{Summary statistics for the time series of Ethereum's technical indicators.}
\label{Table:ethereum_technical_metrics}
\end{table}

From the knowledge of these technical indicators it is possible to calculate the trading indicators. More precisely, we used the \textit{StockStats} Python library to generate them.

We used 36 different trading indicators as shown in Table \ref{tab:trading_indicators}. The lag values represent how previous values ($t-1,\dots,t-n$) are used as input. The {\em window size} indicates the number of previous values used to evaluate the indicator at time $t$, e.g. to calculate $ADXR_t$ at time $t$ we use $ADX_{t-1},...,ADXR_{t-10}$, ten previous values.  

We provide here the definition of the five main trading indicators. 

\begin{itemize}
    \item Simple Moving Average ($SMA$): calculated as the arithmetic average of the cryptocurrency closing price over some period (known as \textit{timeperiod}).
    \item Weighted Moving Average ($WMA$): it is a moving average calculation that assigns higher weights to the most recent price data.
    \item Relative Strength Index ($RSI$): it is a momentum indicator that measures the magnitude of recent price changes. It is normally used to evaluate whether stocks or other assets are being overbought or oversold.
    \item Price Rate Of Change ($ROC$): it measures the percentage change in price between the current price and the price a certain number of periods ago.
    \item Momentum: it is the rate of acceleration of a security's price, i.e. the speed at which the price is changing. This measure is particularly useful to identify trends.
    \item On Balance Volume ($OBV$): it is a technical momentum indicator based on the traded volume of an asset to predict changes in stock price.
\end{itemize}

\begin{table}[!htbp]
\centering
\begin{tabular}{lrrrrrr}
\toprule
{} &         \textbf{SMA} &           \textbf{WMA} &           \textbf{RSI} &          \textbf{ROCP} &           \textbf{MOM} &            \textbf{OBV} \\
\midrule
\textbf{mean} &   7924.974 &   7926.275 &     51.797 &      0.0013 &      8.685 &  126972.751 \\
\textbf{std}&   5465.393 &   5467.642 &     14.484 &      0.027 &    298.311 &   33544.231 \\
\textbf{min}&    767.801 &    766.912 &      2.426 &     -0.321 &  -5260.55 &   18811.069954 \\
\textbf{25\%}&   4135.225 &   4134.525 &     42.374 &     -0.0083 &    -53.64 &  110336.0947 \\
\textbf{50\%}&   7427.187 &   7427.521 &     51.877 &      0.001 &      4.97 &  126464.383 \\
\textbf{75\%}&   9753.094 &   9751.604 &     61.151 &      0.011 &     70.732 &  147814.358 \\
\textbf{max }&  40996.6 &  41106.93 &     98.641 &      0.314 &   4069.26 &  213166.214 \\
\bottomrule
\end{tabular}
\caption{Summary statistics for the time series of Bitcoin's trading indicators.}
\label{Table:bitcoin_technical_trading_metrics}
\end{table}

\begin{table}[!htp]
\begin{tabular}{lrc}
\toprule
\textbf{Trading Indicator}                                                                     & \multicolumn{1}{l}{\textbf{Lag}} & \multicolumn{1}{l}{\textbf{Window size}} \\
\toprule
{ SMA: Simple Moving Average}                    & \multicolumn{1}{c}{-}            & \multicolumn{1}{c}{10}                                        \\
{ WMA: Weighted Moving Average}                    & \multicolumn{1}{c}{-}            & \multicolumn{1}{c}{10}                                        \\
{ RSI: Relative Strength Index}                    & \multicolumn{1}{c}{-}            & \multicolumn{1}{c}{10}                                        \\
{ ROC: Price Rate Of Change}                       & \multicolumn{1}{c}{-}            & \multicolumn{1}{c}{10}                                        \\
{ Mo: Momentum:}                                   & \multicolumn{1}{c}{-}            & \multicolumn{1}{c}{10}                                        \\
{ OBV: On Balance Volume}                          & 1                                & -                                                             \\
{ permutation (zero based)}                        & 1                                & -                                                             \\
{ log return}                                      & 1                                & -                                                             \\
{ max in range}                                    & 1                                & -                                                             \\
{ min in range}                                    & 1                                & -                                                             \\
{ middle = (close + high + low) / 3}               & 1                                & -                                                             \\
{ compare: le, ge, lt, gt, eq, ne}                 & 1                                & -                                                             \\
{ count: both backward(c) and forward(fc)}         & 1                                & -                                                             \\
{ SMA: simple moving average}                      & \multicolumn{1}{c}{-}            & \multicolumn{1}{c}{10}                                        \\
{ EMA: exponential moving average}                 & \multicolumn{1}{c}{-}            & \multicolumn{1}{c}{10}                                        \\
{ MSTD: moving standard deviation}                 & \multicolumn{1}{c}{-}            & \multicolumn{1}{c}{10}                                        \\
{ MVAR: moving variance}                           & \multicolumn{1}{c}{-}            & \multicolumn{1}{c}{10}                                        \\
{ RSV: raw stochastic value}                       & \multicolumn{1}{c}{-}            & \multicolumn{1}{c}{10}                                        \\
{ RSI: relative strength index}                    & \multicolumn{1}{c}{-}            & \multicolumn{1}{c}{10}                                        \\
{ KDJ: Stochastic oscillator}                      & \multicolumn{1}{c}{-}            & \multicolumn{1}{c}{10}                                        \\
{ Bolling: including upper band and lower band.}   & 1                                & -                                                             \\
{ MACD: moving average convergence divergence}     & \multicolumn{1}{c}{-}            & \multicolumn{1}{c}{5}                                         \\
{ CR: price momentum index}                                             & 1                                & -                                                             \\
{ WR: Williams Overbought/Oversold index}          & 1                                & -                                                             \\
{ CCI: Commodity Channel Index}                    & 1                                & -                                                             \\
{ TR: true range}                                  & 1                                & -                                                             \\
{ ATR: average true range}                         & 1                                & -                                                             \\
{ line cross check, cross up or cross down.}       & 1                                & -                                                             \\
{ DMA: Different of Moving Average (10, 50)}       & 1                                & -                                                             \\
{ DMI: Directional Moving Index, including}        & 1                                & -                                                             \\
{ DI: Positive Directional Indicator}              & 1                                & -                                                             \\
{ ADX: Average Directional Movement Index}         & \multicolumn{1}{c}{-}            & \multicolumn{1}{c}{5}                                         \\
{ ADXR: Smoothed Moving Average of ADX}            & \multicolumn{1}{c}{-}            & \multicolumn{1}{c}{10}                                        \\
{ TRIX: Triple Exponential Moving Average}         & \multicolumn{1}{c}{-}            & \multicolumn{1}{c}{10}                                        \\
{ TEMA: Another Triple Exponential Moving Average} & \multicolumn{1}{c}{-}            & \multicolumn{1}{c}{10}                                        \\
{ VR: Volatility Volume Ratio}                     & 1                                & -                                                       \\
\bottomrule
\end{tabular}
\caption{Trading indicators with associated lags and window size. Lags represent how previous values at ($t-1,\dots,t-n$) are used as input. Window size represents the number of previous values used to compute the indicator at time $t$, e.g. to calculate $ADXR_t$ at time $t$ we use $ADX_{t-1},\dots,ADXR_{t-10}$.}
\label{tab:trading_indicators}
\end{table}

\begin{figure}[htbp]
\centering
\includegraphics[width=\textwidth]{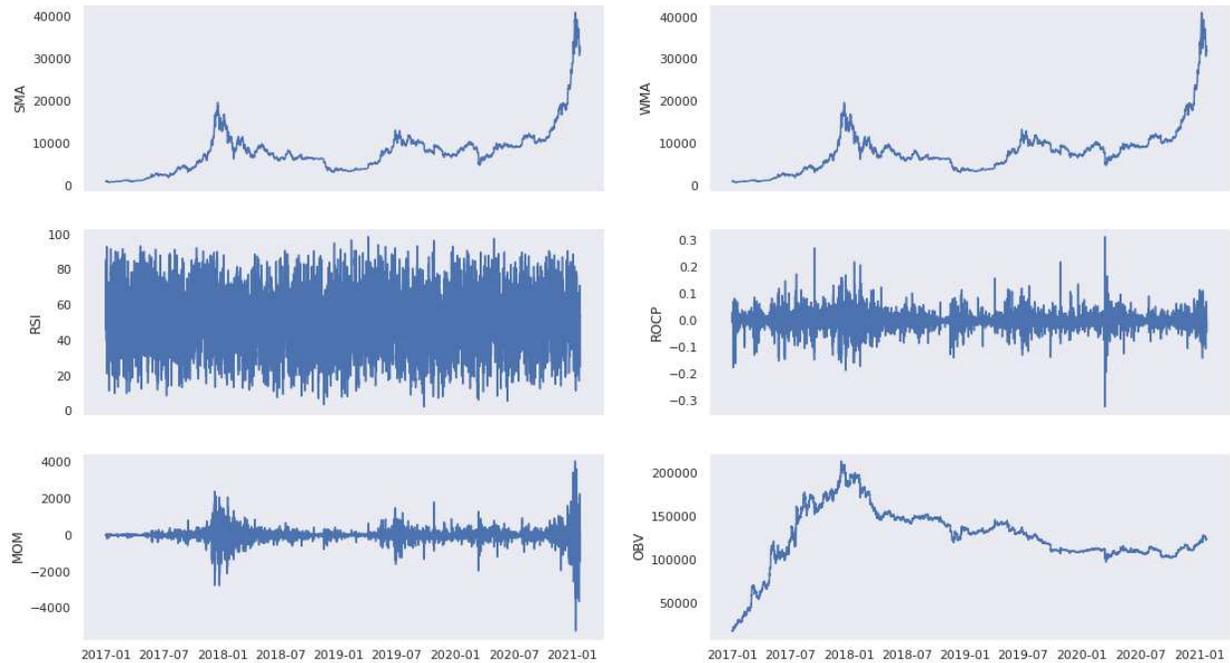}
\caption{Plot of the time series of Bitcoin trading indicators.}
\label{figure:bitcoin_technical_trading_metrics}
\end{figure}

\begin{table}[htbp]
\centering
\begin{tabular}{lrrrrrr}
\toprule
{} &         \textbf{SMA} &           \textbf{WMA} &           \textbf{RSI} &          \textbf{ROCP} &           \textbf{MOM} &            \textbf{OBV} \\
\midrule
\textbf{mean}  & 310.723 &    310.78 &     51.18 &      0.002 &      0.378 &  6.776e+05 \\
\textbf{std}   &    245.768 &    245.87 &     14.246 &      0.035 &     16.607 &  5.739e+05 \\
\textbf{min}   & 8.147 &      8.163 &      3.797 &     -0.317 &   -239.48 & -4.993e+04 \\
\textbf{25\%}  &160.202 &    160.252 &     42.063 &     -0.012 &     -2.8 &  9.864e+04 \\
\textbf{50\%}  &230.916 &    230.962 &     51.038 &      0.000934 &      0.09 &  5.185e+05 \\
\textbf{75\%}  &388.081 &    388.125 &     60.338 &      0.016 &      3.67 &  1.246e+06 \\
\textbf{max}   &1404.89 &   1411.776 &     95.799 &      0.333 &    262.36 &  1.667e+06 \\
\bottomrule
\end{tabular}
\caption{Summary statistics for the time series of Ethereum's trading indicators.}
\label{Table:ethereum_technical_trading_metrics}
\end{table}

\begin{figure}[htbp]
\centering
\includegraphics[width=\textwidth]{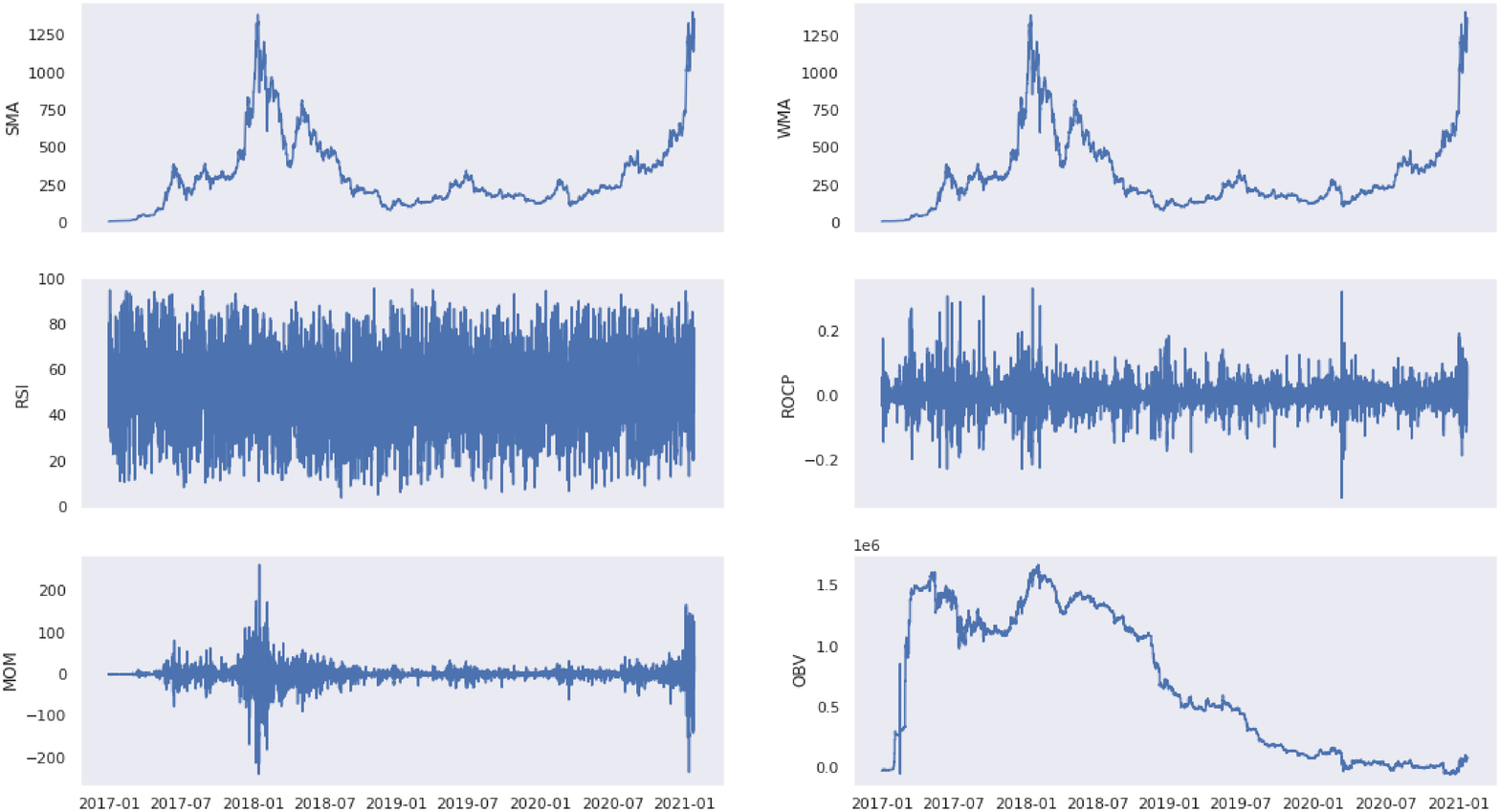}
\caption{Plot of the time Series of Ethereum trading indicators.}
\label{figure:ethreum_technical_trading_metrics}
\end{figure}

Tables \ref{Table:bitcoin_technical_trading_metrics} and \ref{Table:ethereum_technical_trading_metrics} show the statistics of the trading indicators for the considered period of analysis. In Figures \ref{figure:bitcoin_technical_trading_metrics} and \ref{figure:ethreum_technical_trading_metrics} we can see the same trading indicators in a historical time series plot. 
Technical and Trading indicators are used in the next sections to create a model for the prices classification. 

\subsection{Social Media Indicators}
\label{subsec:social_media_indicators}

This section describes how the time series of social media indicators are constructed from Ethereum and Bitcoin developers comments on Github and users' comments on Reddit respectively. In particular, for Reddit we considered the four sub-Reddit channels listed in Table \ref{Table:subreddits}. The time period considered ranges from January 2017 to January 2021.

\begin{table}[!htbp]

\begin{center}
\begin{tabular}{rll}
\toprule
\textbf{\textit{Cryptocurrency}}& \textbf{\textit{Technical Discussions}}& \textbf{\textit{Trading Discussions}} \\
\midrule
Bitcoin & r/Bitcoin & r/BitcoinMarkets \\
Ethereum & r/Ethereum & r/EthTrader \\
\bottomrule
\end{tabular}%
\caption{List of sub-Reddit channels considered in the analysis.}
\label{Table:subreddits}
\end{center}
\end{table}

Examples of a developer's comment extracted from Github for Ethereum and user's comment extracted from Reddit r/Ethereum can be seen in Tables \ref{table:tableGithubEmotionsExample}, \ref{table:tableEmotionsRedditExample}. Quantitative measures of sentiment and emotions associated with the comments, as reported in this example, are computed using state-of-the-art textual analysis tools (further detailed below). These social media indicators computed for each comment are
emotions as love (L), joy (J), anger (A), sadness (S), VAD (valence (Val), dominance (Dom), arousal (Ar)) and sentiment (Sent).

\begin{table}[htbp]
\begin{center}
    \begin{tabular}{|m{4cm}|c|c|c|c|c|c|c|c|}
    \hline
    \textbf{Comment} & 
    \textbf{L} &
    \textbf{J} & 
    \textbf{A} & 
    \textbf{S} & 
    \textbf{Val} & 
    \textbf{Dom} &
    \textbf{Ar} & 
    \textbf{Sent} \\
 \hline
   \textit{\small{Perhaps there's simply nothing new to translate? The reason I updated Transifex in the first place was to be sure the strings with subtle English changes (that don't change the meaning) didn't reset the translation - so those were imported from the old translations. Though I seem to recall at least one truly new string - Transaction or such.}} &
   0 &
   0 &
   0 &
   1 &
   1.93 &
   1.88 &
   1.26 &
   0 \\
 \hline

\end{tabular}
\end{center}
\caption{Example of a Github comment and corresponding emotions (love (L), joy (J), anger (A), sadness (S)), VAD (valence (Val), dominance (Dom), arousal (Ar)), politeness and sentiment (Pol and Sent respectively).}
\label{table:tableGithubEmotionsExample}

\end{table}

\begin{table}[htbp]
\begin{center}    
    \begin{tabular}{|m{4cm}|c|c|c|c|c|c|c|c|}
    \hline
    \textbf{Comment} & 
    \textbf{L} &
    \textbf{J} & 
    \textbf{A} & 
    \textbf{S} & 
    \textbf{Val} & 
    \textbf{Dom} &
    \textbf{Ar} & 
    \textbf{Sent} \\
 \hline
   \textit{\small{All the tosspots focusing on Vitaliks wealth completely miss the point.
If the crypto you are supporting has a purpose it will garner interest in the real world therefore the capital will flow to it. All is measured on the merit and proper fundamentals and not twitterbot pump and dumps...}} &
   0 &
   0 &
   0 &
   0 &
   2.13 &
   1.98 &
   2.26 &
   -1 \\
 \hline

\end{tabular}
\end{center}
\caption{Example of Reddit comment and correspondent emotions (love (L), joy (J), anger (A), sadness (S)), VAD (valence (Val), dominance (Dom), arousal (Ar)), politeness and sentiment (Pol and Sent respectively).}
    \label{table:tableEmotionsRedditExample}
        
\end{table}

\subsection{Social Media Indicators Evaluation Through Deep Learning}
\label{subsec:social_media_indicators_deep_learning}

\begin{figure}[htbp]
\centering
\includegraphics[width=0.3\textwidth]{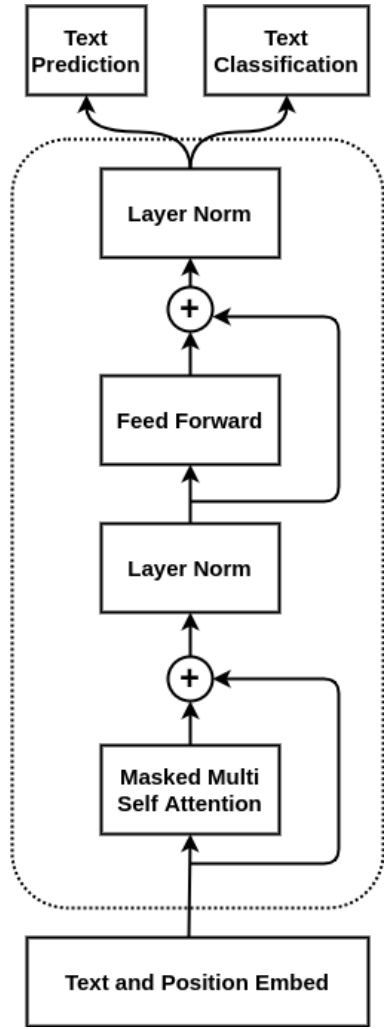}
\caption{ Transformer architectural scheme of Vaswani et al.\cite{vaswani2017attention}. }
\label{figure:transformers_schema}
\end{figure}

We extracted the social media indicators using deep, pre-trained, neural networks called Bidirectional Encoder Representations from Transformers (BERT) \cite{devlin2018bert}.
BERT and other Transformer encoder architectures have been successful in performing various tasks in natural language processing (NLP)  and represent the evolution of Recurrent Neural Network (RNN) typically used in NLP. They compute vector-space representations of natural language that are suitable for use in deep learning models. The BERT family of models uses the Transformer encoder architecture to process each token of input text in the full context of all tokens before and after, hence the name: Bidirectional Encoder Representations from Transformers. BERT models are usually pre-trained on a large corpus of text, then fine-tuned for specific tasks. These models provide dense vector representations for natural language by using a deep, pre-trained neural network with the Transformer architecture represented in Figure \ref{figure:transformers_schema}. 

Transformers are based on the \textit{Attention Mechanism} where RNN units would encode the input up until timestamp $t$ into one hidden vector $h_t$. The latter would then be passed to the next timestamp (or to the decoder in the case of a sequence-to-sequence model). By using the attention mechanism, one no longer tries to encode the full source sentence into a fixed-length vector. Instead, one allows the decoder to attend to different parts of the source sentence at each step of the output generation. Importantly, we let the model learn what to attend to based on the input sentence and what it has produced so far.

\begin{figure}[htbp]
\centering
\includegraphics[width=0.5\textwidth]{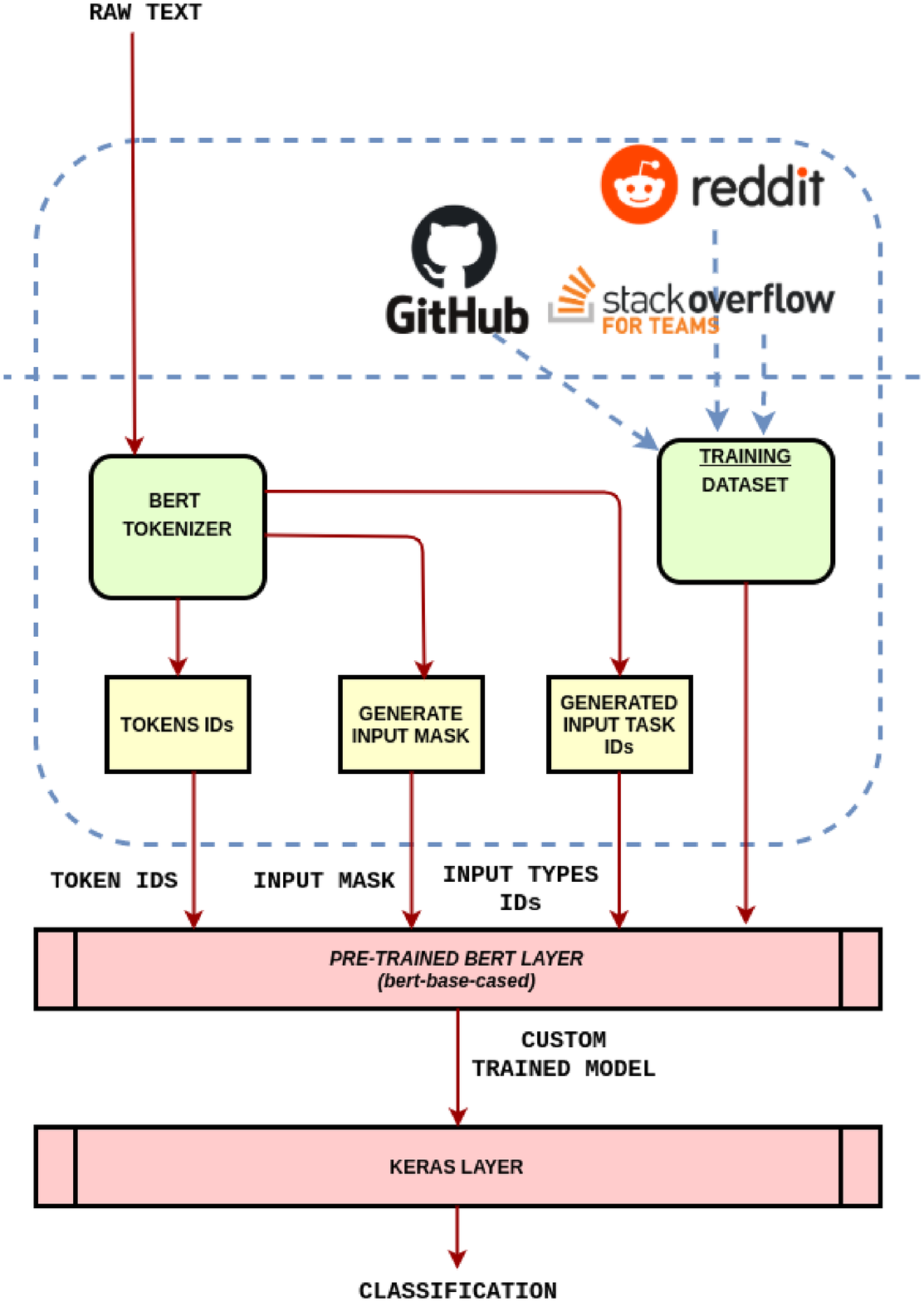}
\caption{Scheme of the general bidirectional encoder representation from Transformer. }
\label{figure:bert_schema}
\end{figure}

The Transformer architecture allows for the creation of NLP models trained on very large datasets as we have done in this work. It is feasible to train such models on large datasets thanks to pre-trained language models, which can be fine-tuned on the particular dataset without the effort of re-training the whole network.

The weights learnt by the extensively pre-trained models can be later reused for specific tasks by simply tailoring the weights to the specific dataset.
This would allow us to exploit what the pre-trained language model has learnt with a finer weight tuning by capturing the lower-level intricacies of the specific dataset.

We used Tensorflow and Keras Python libraries with the Transformer package to leverage the power of these pre-trained neural networks. In particular, we used the \textit{BERT-base-case} pre-trained model. Figure \ref{figure:bert_schema} shows the architectural design used to train the three NN classifiers used to extract the social media indicators. This figure shows the three gold datasets used to train our final models, namely Github, Stack Overflow and Reddit.

In particular, we used a sentiment-labelled dataset consisting of $4423$ posts mined from Stackoverflow user's comments to train the sentiment model for Github: comments on both platforms are written using the technical jargon language of software developers and engineers. We also used an emotion-labelled dataset of $4200$ sentences from Github \cite{murgia2018exploratory}. Finally, we used a sentiment-labelled dataset containing  more than 33K labelled Reddit users's comments\footnote{https://www.kaggle.com/cosmos98/twitter-and-reddit-sentimental-analysis-dataset}.

Tables \ref{tab:reddit_clf}, \ref{tab:github_sentiment_clf} and \ref{tab:github_emotion_clf} show the performance of sentiment and emotion classification on the two different dataset: Github and Reddit.

\begin{table}[!htpb]
\centering
\begin{tabular}{llll}
\toprule
                      & \textbf{precision}          & \textbf{recall}             & \textbf{f1-score}           \\
\midrule
\textbf{negative}     & 0.92                        & 0.89                        & 0.90                        \\
\textbf{neutral}      & 0.97                        & 0.98                        & 0.98                        \\
\textbf{positive}     & 0.95                        & 0.95                        & 0.95                        \\
\textbf{accuracy}     &                             &                             & 0.95                        \\
\textbf{macro avg}    & 0.95                        & 0.94                        & 0.94   
\\
\midrule
\textbf{weighted avg} &  0.95 &  0.95 &  0.95\\
\bottomrule
\end{tabular}
\caption{Sentiment Classifier Evaluation For Reddit.}
\label{tab:reddit_clf}
\end{table}

\begin{table}[!htpb]
\centering
\begin{tabular}{llll}
\toprule
                      & \textbf{precision} & \textbf{recall} & \textbf{f1-score} \\
\midrule
\textbf{negative}     & 0.98               & 0.85            & 0.91              \\
\textbf{neutral}      & 0.84               & 0.94            & 0.89              \\
\textbf{positive}     & 0.96               & 0.97            & 0.96              \\
\textbf{accuracy}     &                    &                 & 0.92              \\
\textbf{macro avg}    & 0.93               & 0.92            & 0.92              \\
\midrule
\textbf{weighted avg} & 0.93               & 0.92            & 0.92             \\
\bottomrule
\end{tabular}
\caption{Sentiment Classifier Evaluation For Github.}
\label{tab:github_sentiment_clf}
\end{table}

\begin{table}[!htpb]
\centering
\begin{tabular}{llll}
\toprule
                      & \textbf{precision} & \textbf{recall} & \textbf{f1-score} \\
\midrule
\textbf{anger}        & 0.83               & 0.77            & 0.80              \\
\textbf{sadness}      & 0.89               & 0.89            & 0.89              \\
\textbf{joy}          & 0.86               & 1.00            & 0.92              \\
\textbf{love}         & 1.00               & 1.00            & 1.00              \\
\textbf{accuracy}     &                    &                 & 0.89              \\
\textbf{macro avg}    & 0.89               & 0.91            & 0.90              \\
\midrule
\textbf{weighted avg} & 0.89               & 0.89            & 0.89              \\
\bottomrule
\end{tabular}
\caption{Emotion Classifier Evaluation For Github.}
\label{tab:github_emotion_clf}
\end{table}

\subsubsection{Social Media Indicators on Github}
\label{subsubsec:affect_metrics_github}

Both the Bitcoin and Ethereum projects are open-source, hence the code and all the interactions among contributors are publicly available on GitHub \cite{ortu2018mining}. 
Active contributors are continuously opening, commenting, and closing the so-called ``issues''. An issue is an element of the development process, which carries information about discovered bugs, suggestions on new functionalities to be implemented in the code, new features, or new functionalities being developed. 
It constitutes an elegant and efficient way of tracking all the development process phases, even in complicated and large-scale projects with a large number of remote developers involved. 
An issue can be ``commented'', meaning that developers can start sub-discussions around it. They usually add comments to a given issue to highlight the actions being undertaken or provide suggestions on the possible resolution.
Each comment posted on GitHub is timestamped; therefore it is possible to obtain the exact time and date and generate a time series for each affect metric considered in this study. 

For emotion detection we use the BERT classifier explained in \ref{subsec:social_media_indicators_deep_learning} trained with the public Github's emotion dataset developed by Ortu et al. \cite{murgia2014developers} and extended by Murgia et al. \cite{murgia2018exploratory}. This dataset is particularly suited for our analysis as the algorithm for emotion detection has been trained on developers' comments extracted from the Jira Issue Tracking System\footnote{ITS are software platform used by open source communities and private software companies to manage the development process.} of the Apache Software Foundation, hence within the Software Engineering domain and context of Github and Reddit (considering the selected subreddits).
The classifier can detect love, anger, joy and sadness with an $F_1$ score\footnote{The $F_1$ score tests the accuracy of a classifier. It is calculated as the harmonic mean of precision and recall.} close to $0.89$ for all of them.
 
Valence, Arousal and Dominance (VAD) represent conceptualised affective dimensions that respectively describe the interest, alertness and control a subject feels in response to a particular stimulus. In the context of software development, VAD measures may indicate the involvement of a developer in a project as well as their confidence and responsiveness in completing tasks. Warriner et al.'s~\cite{Warriner2013} has created a reference lexicon containing 
14,000 English words with VAD scores for Valence, Arousal, and Dominance, that can be used to train the classifier, similarly to the approach by  Mantyla et al. \cite{mantyla2016mining}.  In \cite{mantyla2016mining} they extracted the valence-arousal-dominance (VAD) metrics from 700,000 Jira issue reports containing over 2,000,000 comments and showed that issue reports of different type (e.g., feature request vs bug) had a fair variation of valence. In contrast, an increase in issue priority typically increased arousal.

Finally, sentiment is measured using the BERT classifier explained in \ref{subsec:social_media_indicators_deep_learning} trained with the public dataset used in similar studies \cite{calefato2018journal, calefato2017sentiment}. The algorithm extracts the sentiment polarity expressed in short texts in three levels: positive (1), neutral (0) and negative (-1) sentiment.

Our analysis focuses on three main classes of affect metrics: emotions (love, joy, anger, sadness), VAD (valence, arousal, dominance) and Sentiment. As we specify in Section \ref{subsec:social_media_indicators_deep_learning}, we use a tailor-made tool to extract it from the text of the comments for each affect metric class.

Once numerical values of the affect metrics are computed for all comments (as shown in the example in Tables \ref{table:tableGithubEmotionsExample} and \ref{table:tableEmotionsRedditExample}), we consider the comments timestamps (i.e. dates when the comments was posted) to build the corresponding social media time series. The affect time series are constructed aggregating sentiment and emotions of multiple comments on each hour and day depending on the time resolution considered (hourly and daily).

For a given social media indicator, e.g. \textit{anger}, and for a specific time resolution, we construct the time series by averaging the values of the affect metric over all comments posted on the same day.

\begin{figure}[htbp]
\centering
\includegraphics[width=\textwidth]{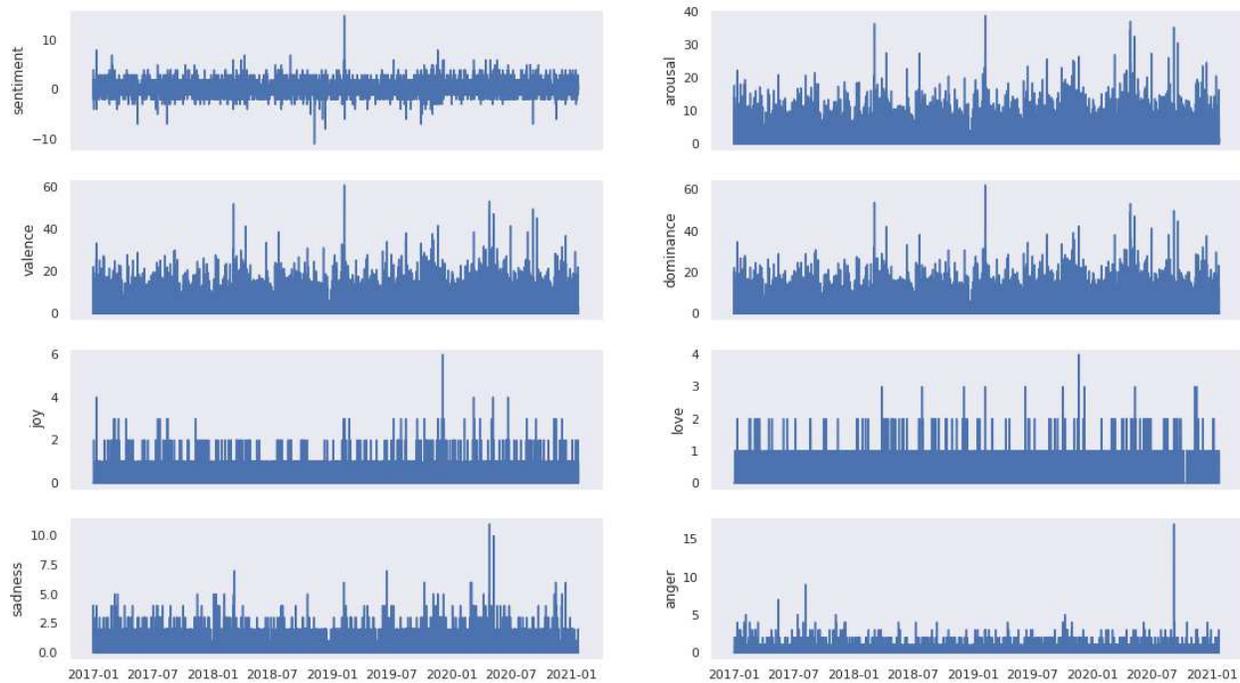}
\caption{Social Media Indicators time series extracted from Github Bitcoin developers comments.}
\label{figure:timeseriesGithubBitcoin}
\end{figure}

\begin{figure}[htbp]
\centering
\includegraphics[width=\textwidth]{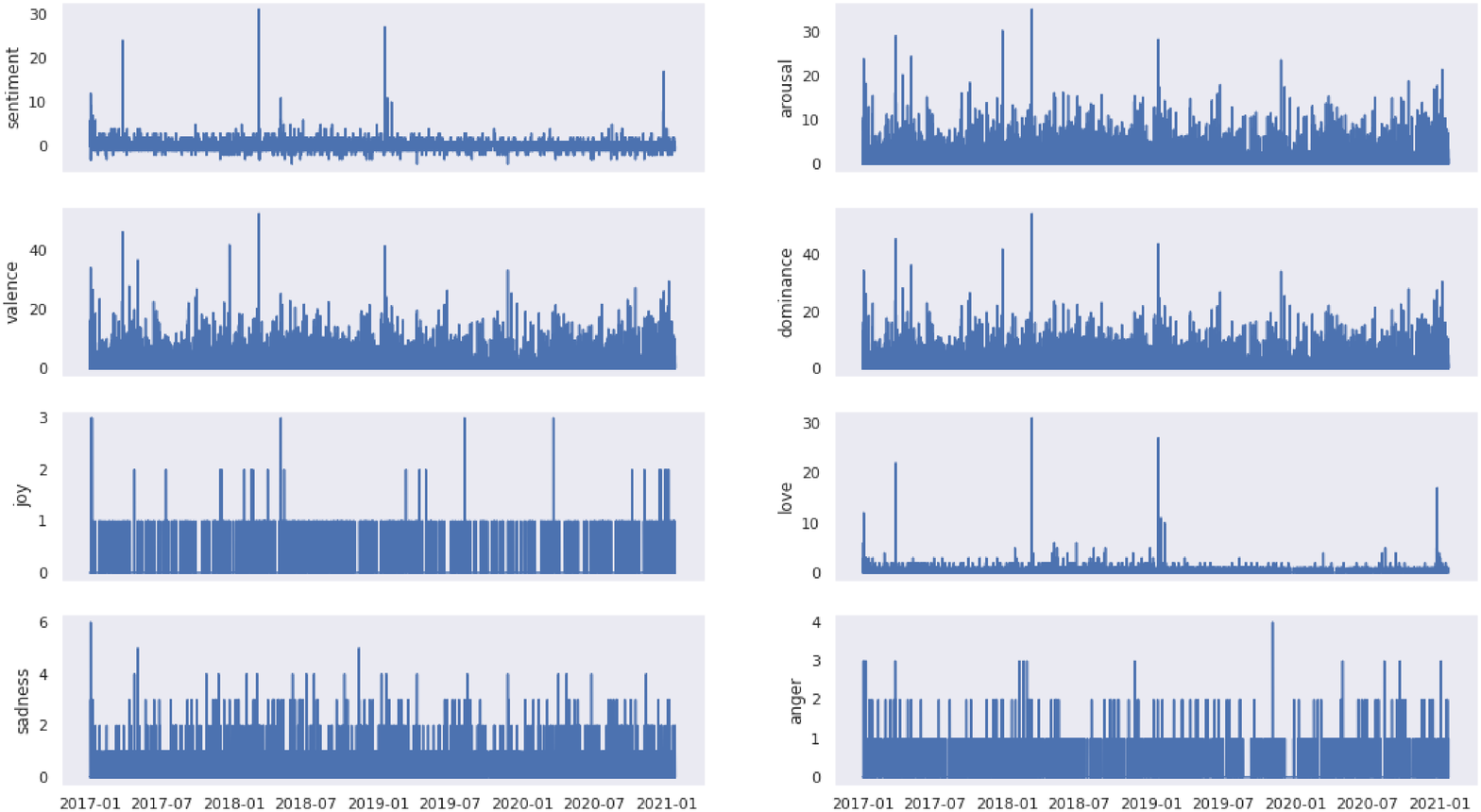}
\caption{Social Media Indicators time series extracted for Github Ethereum developers comments.}
\label{figure:timeseriesGithubEthereum}
\end{figure}

In Table \ref{table:affectMetricsSummaryGithubBitcoin} and \ref{table:affectMetricsSummaryGithubEthereum} we report in more details the summary statistics of the social indicators' time series for both cryptocurrencies respectively. We also report in Figure \ref{figure:timeseriesGithubBitcoin} and \ref{figure:timeseriesGithubEthereum} the time series for all social media indicators for Bitcoin and Ethereum, respectively

\begin{table}[!htpb]
\centering
\begin{minipage}[b]{\linewidth}
\centering
\resizebox{\textwidth}{!}{%
\begin{tabular}{lrrrrrrrr}
\toprule
{} &     \textbf{sentiment} &       \textbf{arousal} &       \textbf{valence} &     \textbf{dominance} &           \textbf{joy} &          \textbf{love} &       \textbf{sadness} &         \textbf{anger} \\
\midrule
\textbf{mean}  &      0.141 &      2.273 &      3.321 &      3.365 &      0.0729 &      0.056 &      0.227 &      0.109 \\
\textbf{std}   &      0.774429 &      2.953897 &      4.324653 &      4.376877 &      0.293435 &      0.248373 &      0.566562 &      0.393479 \\
\textbf{min}   &    -11 &      0 &      0 &      0 &      0 &      0 &      0 &      0 \\
\textbf{25\%}   &      0 &      0 &      0 &      0 &      0 &      0 &      0 &      0 \\
\textbf{50\%}   &      0 &      1.27 &      1.85 &      1.87 &      0 &      0 &      0 &      0 \\
\textbf{75\%}   &      0 &      3.29 &      4.8 &      4.87 &      0 &      0 &      0 &      0 \\
\textbf{max}   &     15 &     38.88 &     60.78 &     62.28 &      6 &      4 &     11 &     17 \\
\bottomrule
\end{tabular}
}
\caption{Summary statistics of Github affect  metrics for Bitcoin.}
\label{table:affectMetricsSummaryGithubBitcoin}
\end{minipage}\hfill
\end{table}

\begin{table}[!htpb]
\centering
\begin{minipage}[b]{\linewidth}
\centering
\resizebox{\textwidth}{!}{%
\begin{tabular}{lrrrrrrrr}
\toprule
{} &     \textbf{sentiment} &       \textbf{arousal} &       \textbf{valence} &    \textbf{dominance} &           \textbf{joy} &          \textbf{love} &       \textbf{sadness} &         \textbf{anger} \\
\midrule
\textbf{mean}  &      0.0842 &      0.7934 &      1.1405 &      1.147 &      0.0182 &      0.0761 &      0.0961 &      0.0356 \\
\textbf{std}   &      0.6754 &      1.7082 &      2.4653 &      2.477 &      0.1407 &      0.5284 &      0.3570 &      0.2052 \\
\textbf{min}   &     -4 &      0 &      0 &      0 &     0 &      0 &      0 &      0 \\
\textbf{25\%}   &     0 &      0 &      0 &      0 &      0 &     0 &      0 &      0 \\
\textbf{50\% }  &      0&      0 &      0&      0 &      0 &      0 &      0 &      0 \\
\textbf{75\% }  &      0 &      1.08&      1.54 &      1.62 &      0 &      0 &      0 &      0 \\
\textbf{max }  &     31&     35.19 &     52.5 &     54.35 &      3 &     31 &      6 &      4 \\
\bottomrule
\end{tabular}
}
\caption{Summary Statistics of Github Social Media Indicators for Ethereum.}
\label{table:affectMetricsSummaryGithubEthereum}
\end{minipage}\hfill
\end{table}

\subsubsection{Measuring Affects Metrics on Reddit}
\label{subsubsec:affect_metrics_reddit}

The social media platform \textit{Reddit} is an American social news aggregation, web content rating, and discussion website that reaches about 8 billion page views per month. It is a top-rated social network in English-speaking countries, especially Canada and the United States. Almost all the messages present are written in English, while the minority, are in Spanish, Italian, French and German.

Reddit is built over multiple subreddits, where each subreddit is dedicated to discussing a particular subject.
Therefore, there are specific subreddits related to major cryptocurrency projects.
For each cryptocurrency in this work, two subreddits are analysed, one technical and one trading related. 
In Tab.~\ref{Table:subreddits} the considered subreddits. are shown.
For each subreddit, we fetched all comments from January 2017 to January 2021.

For emotion detection we use the BERT classifier explained in \ref{subsec:social_media_indicators_deep_learning} trained with the public Github's emotion dataset developed by Ortu et al. \cite{murgia2014developers} and extended by Murgia et al. \cite{murgia2018exploratory}. This dataset is particularly suited for our analysis, as already explained in the previous section.

\begin{table}[!htpb]
\centering
\begin{minipage}[b]{\linewidth}
\centering
\resizebox{\textwidth}{!}{%
\begin{tabular}{lrrrrrrrr}
\toprule
{} &     \textbf{sentiment} &       \textbf{arousal} &       \textbf{valence} &     \textbf{dominance} &           \textbf{joy} &          \textbf{love} &       \textbf{sadness} &         \textbf{anger} \\
\midrule
\textbf{mean}  &      1.8582 &      8.6046&     12.0466 &     11.7412 &      0.6492 &      0.2509 &      0.5579 &      2.2197 \\
\textbf{std}   &      4.3498 &     17.3895 &     24.4038 &     23.7624 &      1.7040 &      0.8278 &      1.4223 &      4.7780 \\
\textbf{min}   &     -9 &      0 &      0&      0 &      0 &      0 &      0 &      0 \\
\textbf{25\% }  &      0 &      0 &      0 &      0 &      0 &      0 &      0 &      0 \\
\textbf{50\%}   &      0 &      1.09 &      1.54 &      1.51 &      0 &      0 &      0 &      0 \\
\textbf{75\%}   &      2 &     10.25 &     14.22 &     13.88 &      1 &      0&      0 &      2 \\
\textbf{max}   &    101 &    492.99 &    680.41 &    662.39 &     42 &     27 &     34&    133 \\
\bottomrule
\end{tabular}
}
\caption{Summary Statistics of Reddit Social Media Indicators for subreddit r/Bitcoin.}
\label{table:affectMetricsSummaryRedditBitcoin}
\end{minipage}\hfill
\end{table}

\begin{figure}[!htbp]
\centering
\includegraphics[width=\textwidth]{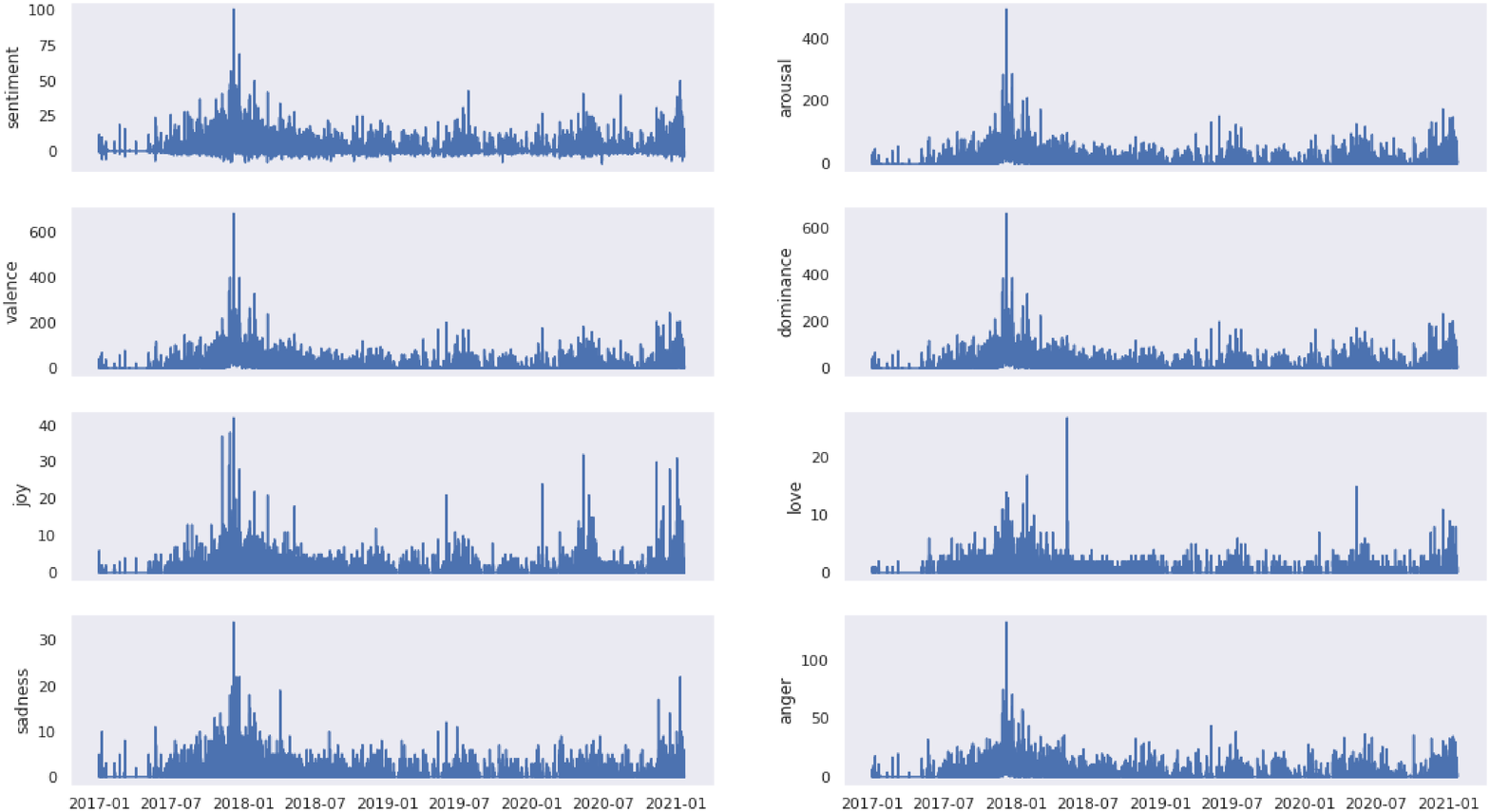}
\caption{Social Media Indicators time series extracted for Reddit  for subreddit r/Bitcoin.}
\label{figure:timeseriesRedditBitcoin}
\end{figure}

\begin{table}[!htpb]
\centering
\begin{minipage}[b]{\linewidth}
\centering
\resizebox{\textwidth}{!}{%
\begin{tabular}{lrrrrrrrr}
\toprule
{} &     \textbf{sentiment} &       \textbf{arousal} &       \textbf{valence} &    \textbf{dominance} &           \textbf{joy} &          \textbf{love} &       \textbf{sadness} &         \textbf{anger} \\
\midrule
\textbf{mean}  &      0.0842 &      0.7934 &      1.1405 &      1.147 &      0.0182 &      0.0761 &      0.0961 &      0.0356 \\
\textbf{std}   &      0.6754 &      1.7082 &      2.4653 &      2.477 &      0.1407 &      0.5284 &      0.3570 &      0.2052 \\
\textbf{min}   &     -4 &      0 &      0 &      0 &     0 &      0 &      0 &      0 \\
\textbf{25\%}   &     0 &      0 &      0 &      0 &      0 &     0 &      0 &      0 \\
\textbf{50\% }  &      0&      0 &      0&      0 &      0 &      0 &      0 &      0 \\
\textbf{75\% }  &      0 &      1.08&      1.54 &      1.62 &      0 &      0 &      0 &      0 \\
\textbf{max }  &     31&     35.19 &     52.5 &     54.35 &      3 &     31 &      6 &      4 \\
\bottomrule
\end{tabular}
}
\caption{Summary Statistics of Reddit Social Media Indicators for subreddit r/Ethereum.}
\label{table:affectMetricsSummaryRedditEthereum}
\end{minipage}\hfill
\end{table}

\begin{figure}[!htbp]
\centering
\includegraphics[width=\textwidth]{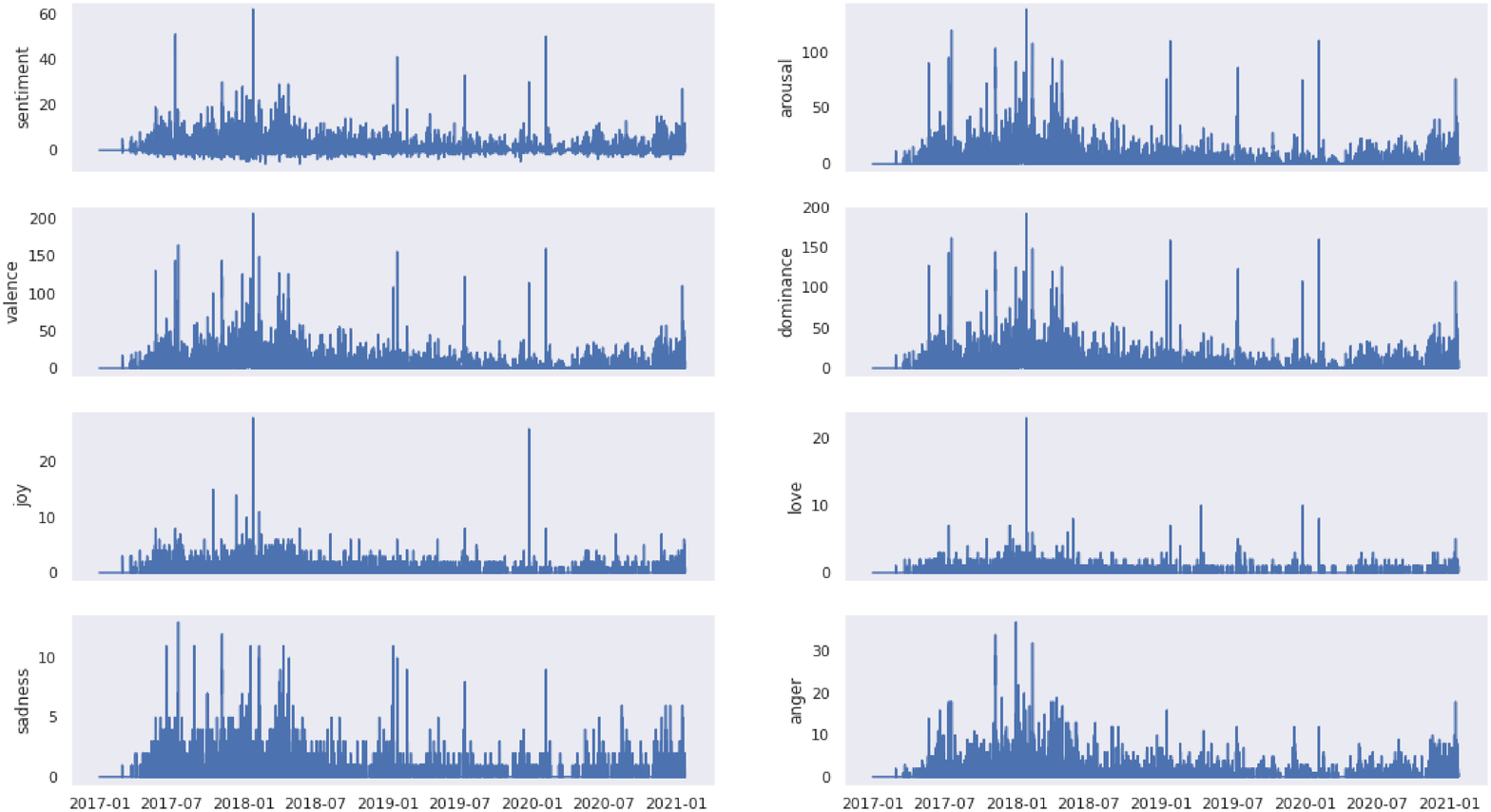}
\caption{Social Media Indicators time series extracted for Reddit for subreddit r/Ethereum.}
\label{figure:timeseriesRedditEthereum}
\end{figure}

\begin{table}[!htpb]
\centering
\begin{minipage}[b]{\linewidth}
\centering
\resizebox{\textwidth}{!}{%
\begin{tabular}{lrrrrrrrr}
\toprule
{} &     \textbf{sentiment} &       \textbf{arousal} &       \textbf{valence} &     \textbf{dominance} &           \textbf{joy} &          \textbf{love} &       \textbf{sadness} &         \textbf{anger} \\
\midrule
\textbf{mean}  &      0.9264 &      4.0327 &      5.6617 &      5.5688 &      0.2419 &      0.1059 &      0.3383 &      1.0095 \\
\textbf{std}   &      2.7650 &     10.7106 &     14.9023 &     14.6243 &      0.8616 &      0.4553 &      1.0275 &      2.8993 \\
\textbf{min}   &     -9 &      0 &      0 &     0 &      0 &      0 &      0 &      0 \\
\textbf{25\%}   &      0 &      0 &      0 &      0 &      0 &      0 &      0 &      0 \\
\textbf{50\%}   &      0 &      0 &      0 &      0 &      0 &      0 &     0 &      0 \\
\textbf{75\%}   &      0&      2.32 &      3.2925 &      3.26&      0 &      0 &      0 &      0 \\
\textbf{max}   &     52 &    245.32 &    332.84 &    329.4 &     34 &     15 &     22 &     88 \\
\bottomrule
\end{tabular}
}
\caption{Summary Statistics of Reddit Social Media Indicators for subreddit r/Bitcoinmakets.}
\label{table:affectMetricsSummaryBitcoinmakets}
\end{minipage}\hfill
\end{table}

\begin{figure}[!htbp]
\centering
\includegraphics[width=\textwidth]{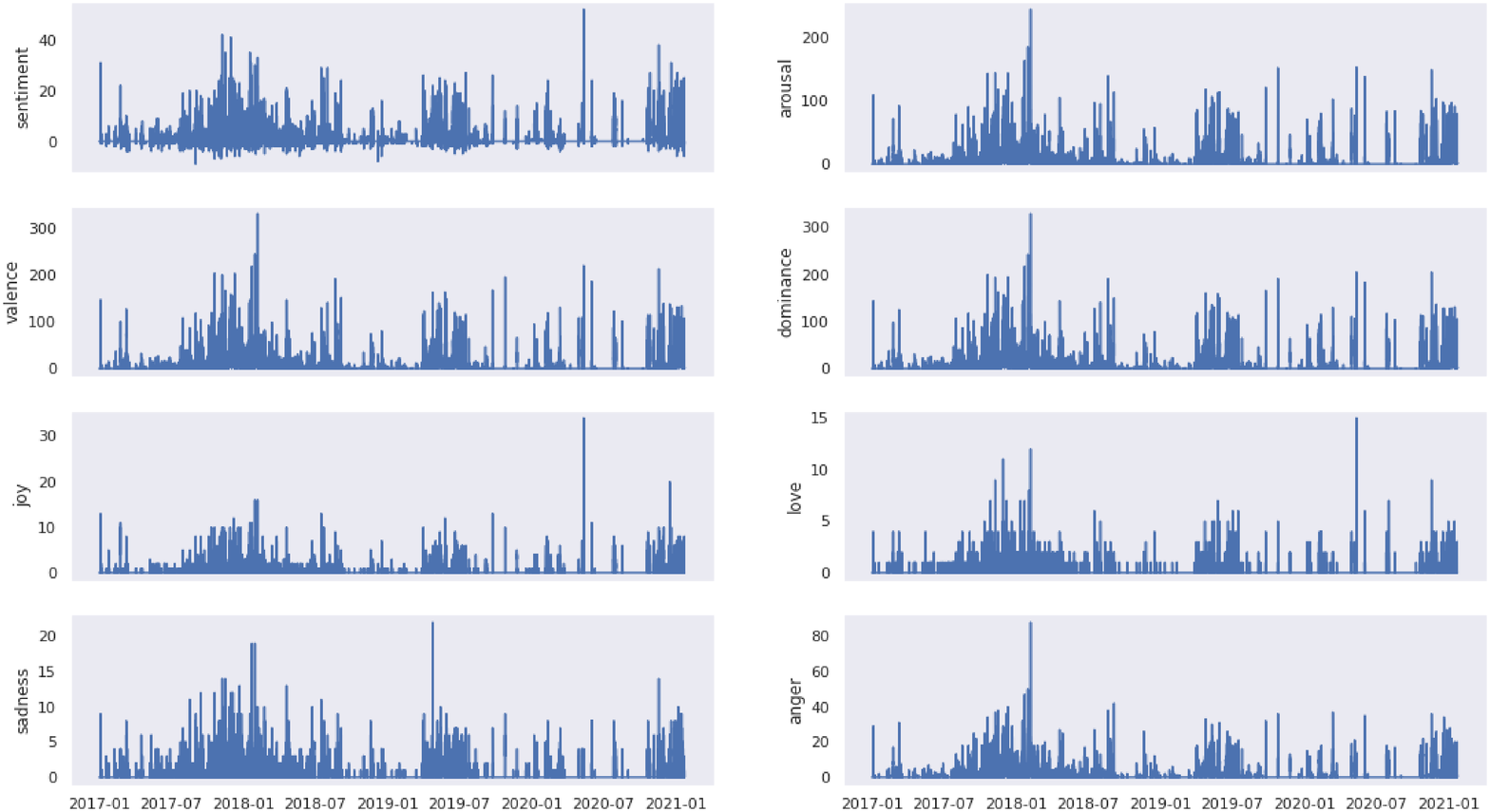}
\caption{Social Media Indicators time series extracted for Reddit for subreddit r/Bitcoinmakets.}
\label{figure:timeseriesRedditBitcoinmakets}
\end{figure}

\begin{table}[!htpb]
\centering
\begin{minipage}[b]{\linewidth}
\centering
\resizebox{\textwidth}{!}{%
\begin{tabular}{lrrrrrrrr}
\toprule
{} &     \textbf{sentiment} &       \textbf{arousal} &       \textbf{valence} &     \textbf{dominance} &           \textbf{joy} &          \textbf{love} &       \textbf{sadness} &         \textbf{anger} \\
\midrule
\textbf{mean}  &      0.8150 &      2.8479 &      4.0716 &      4.0046 &      0.2123 &      0.0855 &      0.2072 &      0.5983 \\
\textbf{std}   &      2.1528 &      6.1395 &      8.7652 &      8.6220 &      0.6807 &      0.3959 &      0.6641 &      1.577632 \\
\textbf{min}   &     -6 &      0 &      0 &      0 &      0 &      0 &      0 &      0 \\
\textbf{25\%}   &      0 &      0 &      0 &      0 &      0&      0 &      0 &      0 \\
\textbf{50\%}   &      0 &      0&      0 &      0 &      0 &      0 &      0 &      0 \\
\textbf{75\%}   &      1 &      3.25&      4.65&      4.6 &      0 &      0 &      0 &      1 \\
\textbf{max}   &     62&    138.39 &    207.38 &    191.95 &     28 &     23 &     13 &     37 \\
\bottomrule
\end{tabular}
}
\caption{Summary Statistics of Reddit Social Media Indicators for subreddit r/Ethtraders.}
\label{table:affectMetricsSummaryRedditEthtraders}
\end{minipage}\hfill
\end{table}

\begin{figure}[!htbp]
\centering
\includegraphics[width=\textwidth]{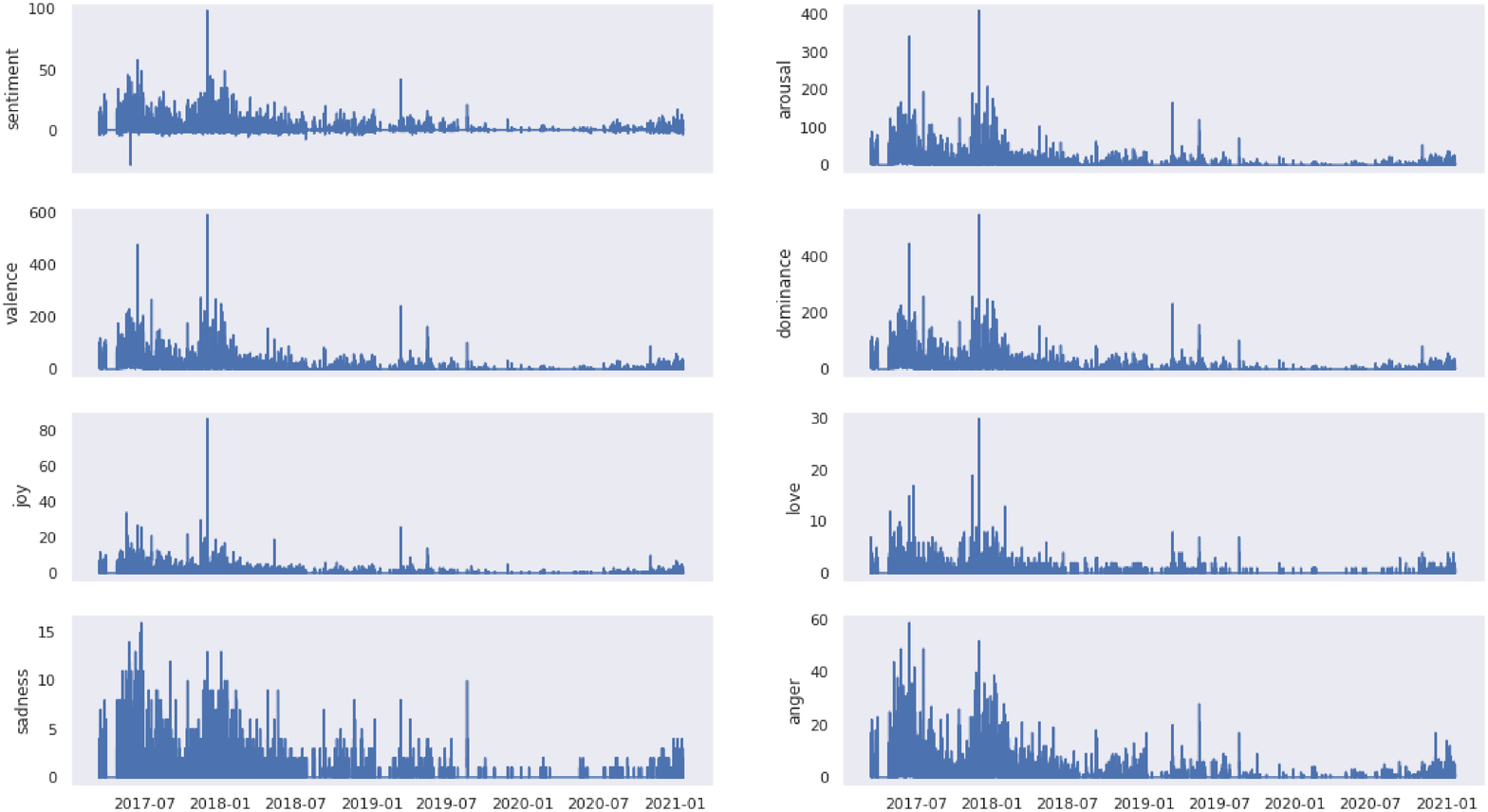}
\caption{Social Media Indicators time series extracted for Reddit for subreddit r/Ethtraders.}
\label{figure:timeseriesRedditETHtraders}
\end{figure}

The classifier can detect love, anger, joy and sadness with an $F_1$ score\footnote{The $F_1$ score tests the accuracy of a classifier and it is calculated as the harmonic mean of precision and recall.} close to $0.89$ for all of them. For VAD metrics we used the same approach in \ref{subsubsec:affect_metrics_github} while for sentiment we used previous approach with BERT deep learning algorithm trained with a public golden  dataset for Reddit comments available in the biggest and well known web platform for sharing datasets Kaggle.com\footnote{https://www.kaggle.com/cosmos98/twitter-and-reddit-sentimental-analysis-dataset}. 

Tables \ref{table:affectMetricsSummaryRedditBitcoin} and \ref{table:affectMetricsSummaryBitcoinmakets} and Figures \ref{figure:timeseriesRedditBitcoin} and \ref{figure:timeseriesRedditBitcoinmakets} show statistics and time series for the two Bitcoin's subreddits  while Tables \ref{table:affectMetricsSummaryRedditEthtraders} and \ref{table:affectMetricsSummaryRedditEthereum} and Figures \ref{figure:timeseriesRedditEthereum} and \ref{figure:timeseriesRedditETHtraders} show statistics and time series for the two Ethereum's subreddits.

\subsection{Price Movement Classification}
\label{sub:price_movement_classification}

The target variable is a binary variable with two unique classes listed below.

\begin{itemize}
    \item \textbf{Upward movements}: This class, labeled with \textit{up} and encoded with $1$, represents the condition of increasing prices.
    \item \textbf{Downward movements}: This class, labeled \textit{down} and encoded with $0$, represents the condition of falling prices.
\end{itemize}

\begin{figure}[!htbp]
    \centering
    \begin{subfigure}{0.48\textwidth}
       \includegraphics[width=\textwidth]{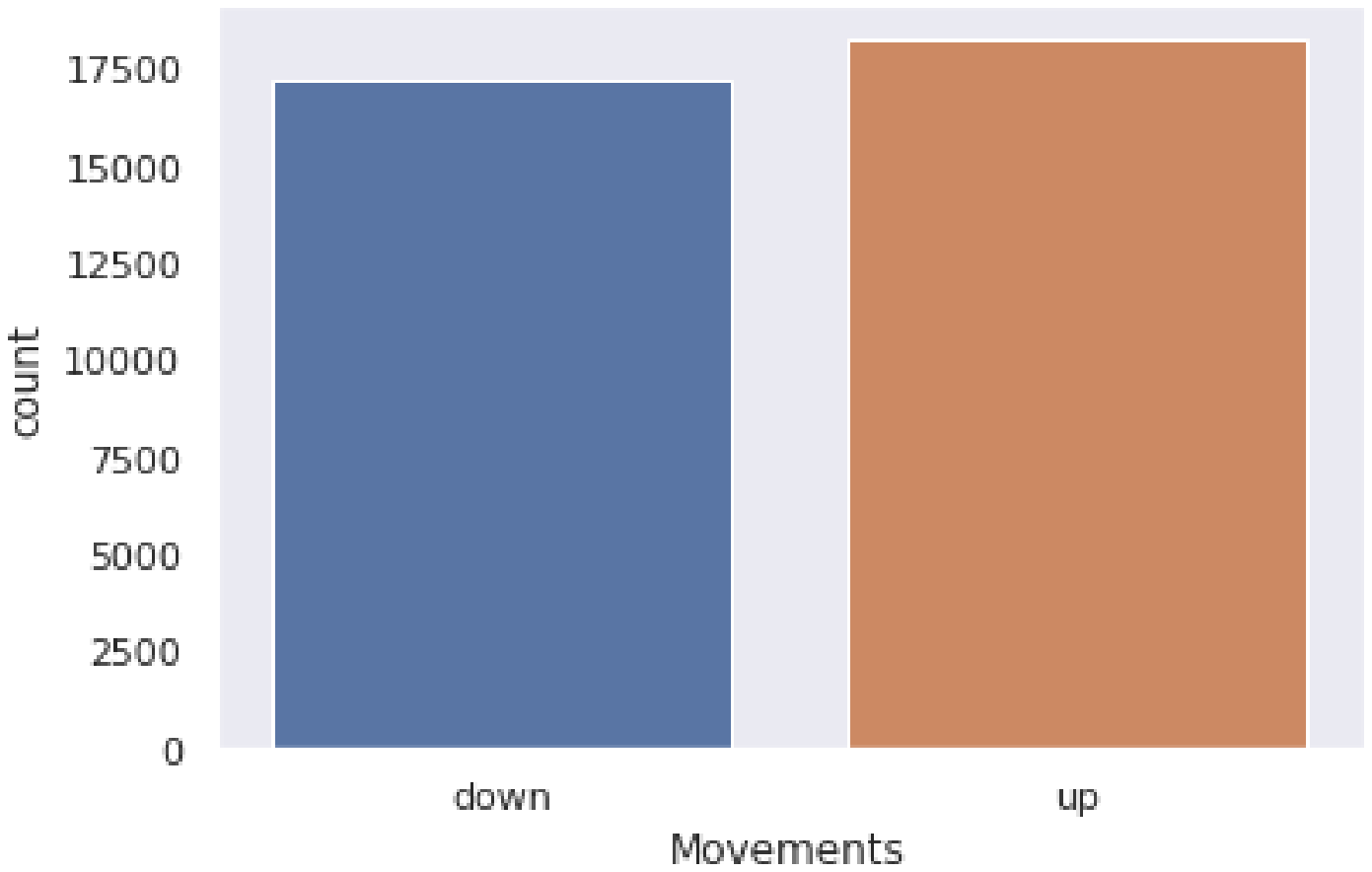}
        \caption{Bitcoin Hourly}
        \label{fig:btc_hourly}
    \end{subfigure}\hspace{1em}%
    \begin{subfigure}{0.48\textwidth}
        \includegraphics[width=\textwidth]{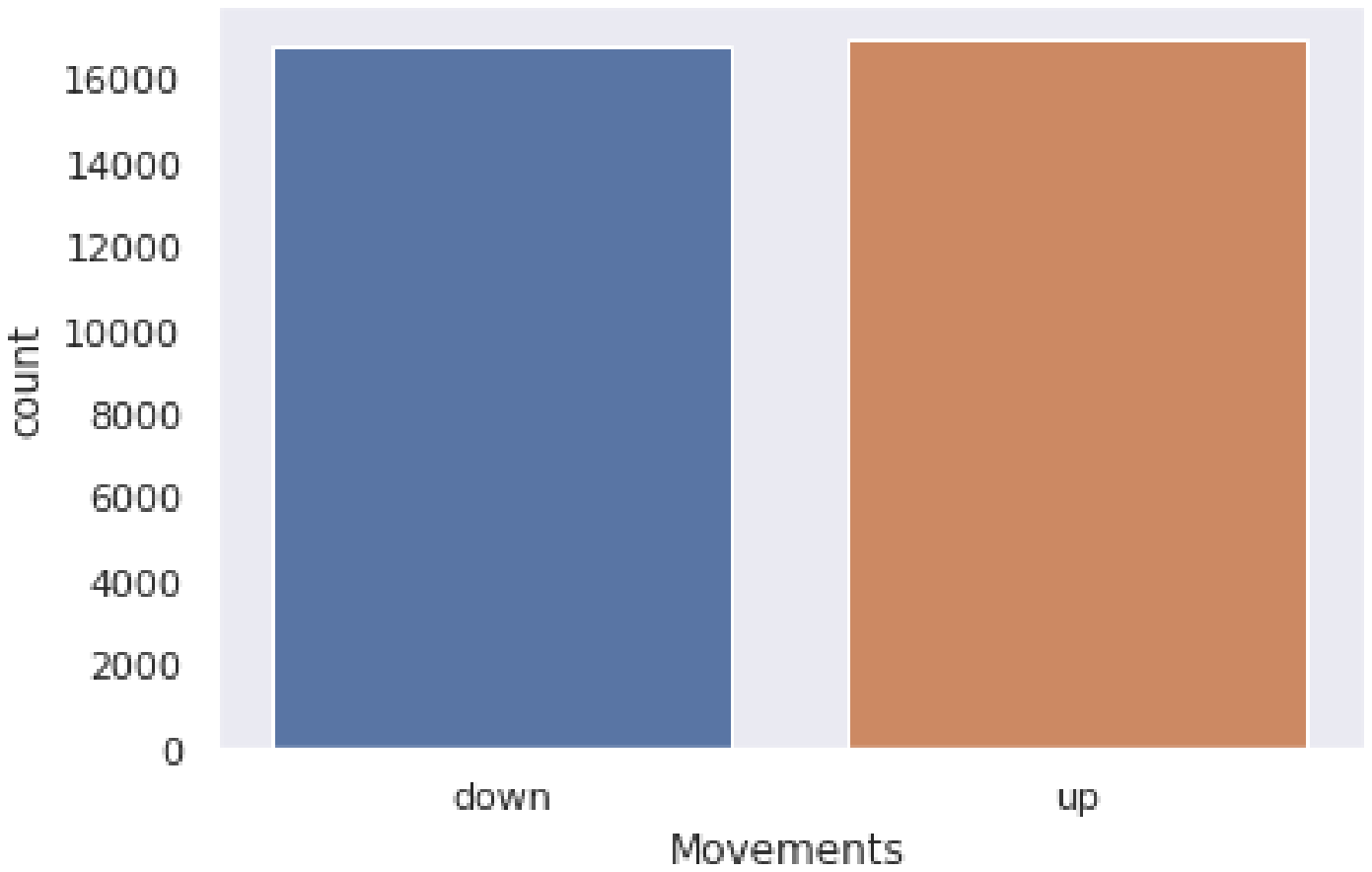}
        \caption{Ethereum Hourly}
        \label{fig:eth_hourly}
    \end{subfigure}
    \begin{subfigure}{0.48\textwidth}
       \includegraphics[width=\textwidth]{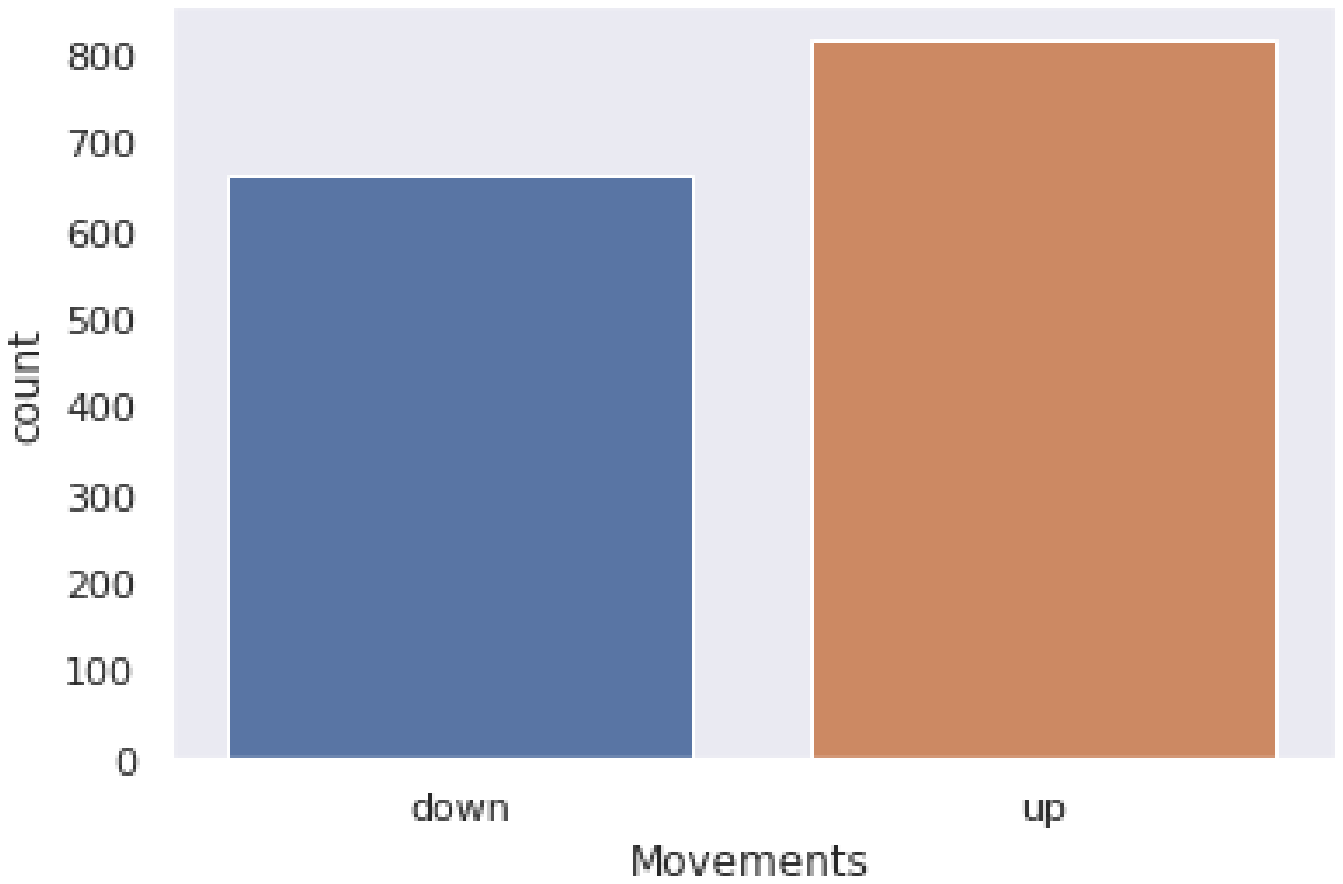}
        \caption{Bitcoin Daily}
        \label{fig:btc_daily}
    \end{subfigure}\hspace{1em}%
    \begin{subfigure}{0.48\textwidth}
        \includegraphics[width=\textwidth]{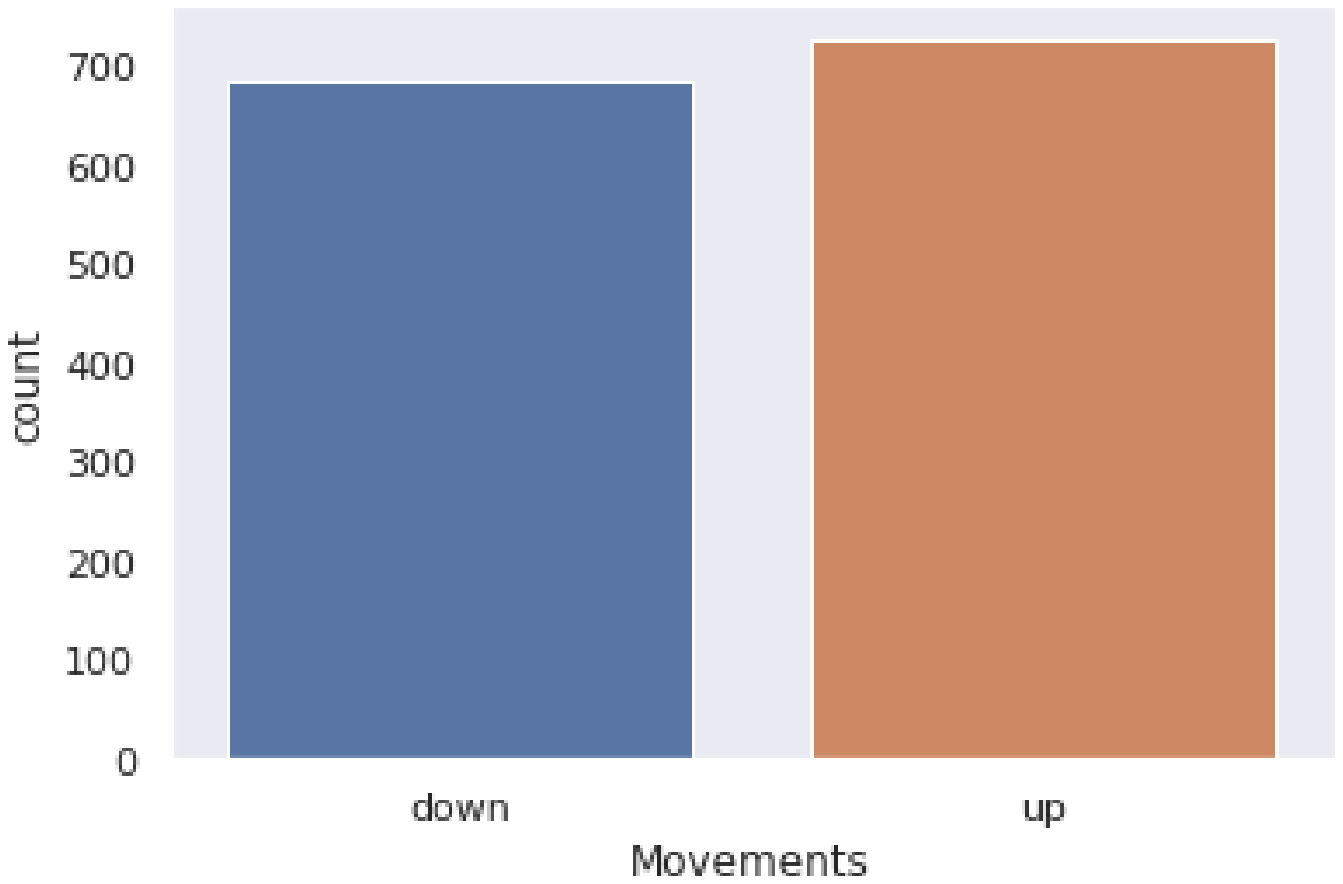}
        \caption{Ethereum Daily}
        \label{fig:eth_daily}
    \end{subfigure}
\caption{Hourly (\ref{fig:btc_hourly},\ref{fig:eth_hourly}) and Daily (\ref{fig:btc_daily},\ref{fig:eth_daily}) Price Movement Classes Distribution.}
\label{fig:MulticlassCountplot}
\end{figure}

Figure \ref{fig:MulticlassCountplot} shows the class distribution and the dataset for hourly and daily frequency, highlighting that we are dealing with fairly balanced classification problems in the case of hourly frequency and slightly unbalanced in the daily frequency case.

\begin{table}[htbp!]
\begin{center}
\begin{tabular}{cclrl}
\toprule
\multicolumn{1}{l}{Frequency} & \multicolumn{1}{l}{Cryptocurrency} & Class & \multicolumn{1}{l}{Counts} & Percentage \\
\toprule
\multirow{4}{*}{Hourly}       & \multirow{2}{*}{Bitcoin}           & up    & 17246                      & 48,5\%     \\
                              &                                    & down  & 18271                      & 51,5\%     \\
                              & \multirow{2}{*}{\textit{Ethereum}} & up    & 16844                      & 49,8\%     \\
                              &                                    & down  & 16956                      & 50,2\%     \\
\multirow{4}{*}{Daily}        & \multirow{2}{*}{Bitcoin}           & up    & 665                        & 44,8\%     \\
                              &                                    & down  & 817                        & 55,8\%     \\
                              & \multirow{2}{*}{\textit{Ethereum}} & up    & 684                        & 48,5\%     \\
                              &                                    & down  & 727                        & 51,5\%    \\
\bottomrule
\end{tabular}
\end{center}
\caption{Class instances counts and percentages for Bitcoin and Ethereum at an hourly or daily frequency.}
\label{table:percentuali}
\end{table}

Table \ref{table:percentuali} shows the details about the instances of classes $down$ and $up$, with $48,5\%$ and $51.5\%$ respectively for Bitcoin and $49,8\%$ and $50,2\%$ for Ethereum with and hourly frequency. For daily frequency we have $44,8\%$ and $55.2\%$  for Bitcoin and $48,5\%$ and $51,5\%$ for Ethereum of $down$ and $up$ class instances. For Bitcoin daily frequency we have a slightly unbalanced distribution toward $up$ classes, in this case we will consider \textit{f1-score} along with \textit{accuracy} to asses the model performance.  

\subsection{Time Series Processing}

Since we are using a supervised learning problem, we prepare our data to have a vector of $x$ inputs and an $y$ output with temporal dependence. In this case, the input vector $x$ is called \textit{regressor}. The $x$ inputs include the model's predictors, i.e. one or several values from the past, the so-called \textit{lagged} values.
Inputs correspond to the values of the selected features discussed in the previous sections. The target variable $y$ is a binary variable, which can be either $0$ or $1$. The $0$ (\textit{down}) instance represents downward price movements. A $0$ instance at time $t$ is obtained when the difference between the \textit{close price} at time $t$ and the \textit{open price} at time $t+1$ is less than or equal to 0. The $1$ (\textit{up}) instance represents upward price movements, i.e. a rising price condition. A $1$ instance is obtained when the difference between the \textit{close price} at time $t$ and the \textit{open price} at next time step $t+1$ is greater than $0$.
We considered two time series models: 
\begin{itemize}
    \item \textit{\textbf{Restricted}}: the input vector $x$ consists of only technical indicators (open, close, high, low, volume).
    \item \textit{\textbf{Unrestricted}}: the input vector $x$ consists of technical, trading and social indicators.
\end{itemize}

For both the \textit{restricted} and \textit{unrestricted} model we used $1$ lagged value for each indicator. 
The purpose of this distinction is to ascertain and quantify whether the addition of trading and social indicators to the \textit{regressor} vector leads to an effective improvement in the Bitcoin and Ethereum price changes classification.

\section{Methodology}
\label{sec:methodology}

This section describes the deep learning algorithms considered in our analysis, followed by a discussion on the fine tuning of the hyper-parameters.

\subsection{Multilayer Perceptron}

A multilayer perceptron (MLP) is a class of feed-forward artificial neural networks (ANNs), characterised by multiple layers of perceptrons and a typical activation function. 

\begin{figure}[!htbp]
\centering
\includegraphics[width=0.7\textwidth]{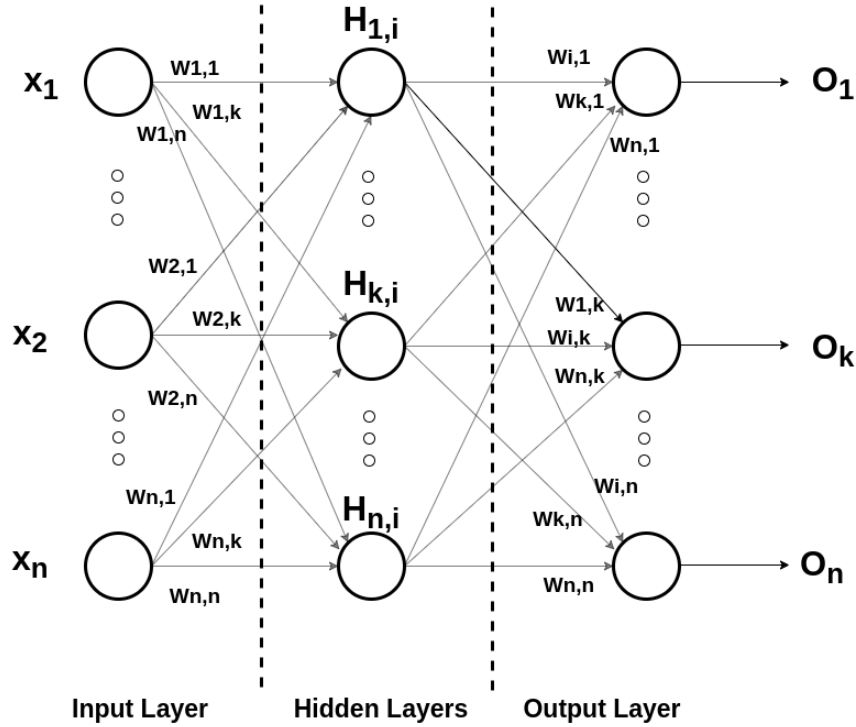}
\caption{Scheme of the Multilayer Perceptron architecture.}
\label{fig:MLP}
\end{figure}

The most common  activation function are:

\begin{equation}
    y(v_i) = \tanh(v_i) ~~ \textrm{and} ~~ y(v_i) = (1+{\rm e}^{-v_i})^{-1} \ ,
    \label{eq:activation_MLP}
\end{equation}

where $v_i$ is the weighted vector of inputs.

When composed of a single hidden layer as in Figure \ref{fig:MLP}, MLPs are called ``vanilla" neural networks (in jargon and for practical use). In general, MLPs refer to neural network architectures with two or more hidden layers.

A MLP comprises three main node categories: input layer nodes, hidden layer nodes and output layer nodes. All nodes of the neural network are perceptrons that use a nonlinear activation function, except for the input nodes. 
MLP differs from a linear perceptron because of its multiple layers and nonlinear activation functions. 

In general, MLP Neural networks are resilient to noise and can also support learning and inference when values are missing. Neural networks do not make strong assumptions about the mapping function and readily learn both linear and nonlinear relationships.
An arbitrary number of input features can be specified, providing direct support for multidimensional forecasting.
An arbitrary number of output values can be specified, providing direct support for multi-step and even multivariate forecasting.
For these reasons, MLP neural networks may be particularly useful for time series forecasting.

In recent developments of deep learning techniques, the rectifier linear unit (ReLU), a piecewise linear function, is frequently used to overcome numerical problems associated with sigmoid functions. Examples of ReLU are the hyperbolic tangent varing between -1 and 1, or the logistic function between 0 and 1. The output of the $i$-th node (neuron) here is $y_{i}$, and the weighted sum of the input connections is $v_{i}$. 

By including the rectifier and softmax functions, alternative activation functions have been developed. Radial basis functions include more advanced activation functions (used in radial basis networks, another class of supervised neural network models).

Since MLPs are fully connected architectures, each node in one layer connects with a specific weight $w_{i,j}$ to every node in the following layer. The neural network is trained using a supervised method called back-propagation and an optimiser method (the Stochastic Gradient Descent is the first and widely used method). After data is processed, learning occurs in the perceptron by adjusting the connection weights, depending on the amount of error in the output relative to the expected result. Back-propagation in the perceptron is a generalisation of the least mean squares (LMS) algorithm.

When the $n_{th}$ training sample is presented to the input layer, the amount of error in the output node $j$ is $e_{j}(n)=d_{j}(n)-y_{j}(n)$, where $d$ is the predicted value and $y$ is the actual value that the perceptron should generate. The back-propagation method then adjusts the node weights to minimise the entire output error provided by Eq. \eqref{eq:learning_1_MLP}:

\begin{equation}
    \begin{array}{l}
    \epsilon(n)=\frac{1}{2}\sum_j e_j^2(n) \ .
    \end{array}
\label{eq:learning_1_MLP}
\end{equation}    

The adjustment of each node's weight is further computed using the gradient descent in Eq. \eqref{eq:learning_2_MLP}, where $y_{i}$ is the output of the previous neuron and $\eta$ is the learning rate:

\begin{equation}
    \begin{array}{l}
    \Delta w_{j,i} (n) = -\eta\frac{\partial\epsilon(n)}{\partial v_j(n)} y_i(n) \ .
    \end{array}
\label{eq:learning_2_MLP}
\end{equation}  

The parameter $\eta$ is commonly set as a trade-off between the weights' convergence to a response and the oscillations around the response.

The induced local field $v_{j}$ varies and one can compute its derivative:
\begin{equation}
    \begin{array}{l}
    -\frac{\partial\epsilon(n)}{\partial v_j(n)} = e_j(n)\phi^\prime (v_j(n))\\
    -\frac{\partial\epsilon(n)}{\partial v_j(n)} = \phi^\prime (v_j(n))\sum_k -\frac{\partial\epsilon(n)}{\partial v_k(n)} w_{k,j}(n) \ ,
    \end{array}
\label{eq:learning_3_MLP}
\end{equation}  
where $\phi^\prime$ is the derivative of the activation function described above, which itself does not vary. 
The analysis is more difficult when modifying the weights of a hidden node, but it can be shown that the relevant quantity is the one showed in Eq. \eqref{eq:learning_3_MLP}.
This algorithm represents a back-propagation of the activation function as Eq. \eqref{eq:learning_3_MLP} depends on the adjustment of the weights of the $k_{th}$ layer, which represent the output layer and this adjustment in turn changes depending on the derivative of the activation functions of the hidden layer weights.

\subsection{Long Short Term Memory}

Long Short-Term Memory networks are a specialised version of Recurrent Neural Network (RNN) able to capture long-term dependencies in a sequence of data.
RNNs are a type of artificial neural networks with a particular topology specialised in the identification of patterns in different types of data sequences: natural language, DNA sequences, handwriting, word sequences, or numerical time series data streams from sensors and financial markets  \cite{hochreiter1997long} for example. Classical recurrent neural networks have a significant disadvantage related to their inability to address long sequences and capture long-term dependencies. RNNs could instead be used only for short sequences with short-term memory dependencies. LSTM were introduced to address the long-term memory problem and are derived directly from RNN to capture long-term dependencies.
An LSTM neural network is organised in units called cells, performing transformations of the input sequence transformation by applying a series operations. An internal state variable is retained by an LSTM cell when forwarded from one cell to the next and is updated by the so-called Operation Gates (forget gate, input gate, output gate) as shown in Figure \ref{fig:AMNN}.  All three gates have different and independent weights and biases, so the network can learn how much of the previous output and current input to maintain and how much of the internal state to pass to the output. Such gates control how much of the internal state is transmitted to the output and operate similarly to other gates.
An LSTM cell unit consists of: 
\begin{enumerate}
    \item A \textit{cell state}:  this state brings information along the entire sequence and represents the  memory of the network.
    \item A \textit{forget gate}:  it filters the relevant information to be kept from previous time steps.
    \item An \textit{input gate}:  it decides what information is relevant to be added add from the current time step.
    \item An \textit{output gate}: it controls the amount of  output at the current time step.
\end{enumerate}

\begin{figure}[htbp]
\centering
\includegraphics[width=0.7\textwidth]{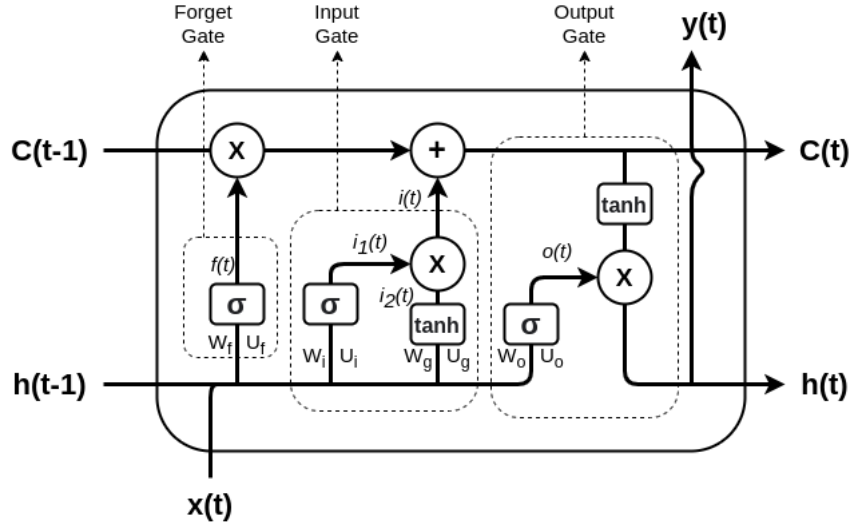}
\caption{LSTM Cell Gate.}
\label{fig:AMNN}
\end{figure}

The first step is the forget gate. This gate takes as input past or lagged values and decides how much of the past information should be forgotten and how much should be saved. The input from the previous hidden state and the current input are transferred through the sigmoid function to the output gate. An output is close to $0$ when that piece of information can be forgotten, while it is close to 1 when that piece of information is to be saved, as follows:

\begin{equation}
    f(t) = \sigma(x(t)*U_f + h(t+1) * W_f)\ .
    \label{eq:f_t_LSTM}
\end{equation}

The matrices $W_{f}$ and $U_f$ contain, respectively, the weights of the input and recurrent connections. The subscript $f$ can be either indicate the forget gate. $x_t$ represents the input vector to the LSTM and $h_{t+1}$ the hidden state vector or output vector of the LSTM unit.

The second gate is the input gate. At this stage the cell state is updated. The previous hidden state and the current input are initially presented as inputs to a sigmoid activation function (the closer the value is to $1$, the more relevant the input is). To boost the network-tuning, it also passes the hidden state and current input to the $\tanh$ function to compress values between $-1$ and $1$. Then the output of the $\tanh$ and of the sigmoid are multiplied element by element (in the formula below the symbol $*$ indicates the multiplication element by element of two matrices). The sigmoid output, in Equation \ref{eq:input_gate_LSTM} determines the information that is important to keep from the $\tanh$ output:

\begin{equation}
    \begin{array}{l}
    i_1(t) = \sigma(x(t)*U_i + h(t+1) * W_i)\ ,\\
    i_2(t) = \tanh(x(t)*U_g + h(t+1) * W_g) \ ,\\
    i(t) = i_1(t)*i_2(t) \ .
    \end{array}
    \label{eq:input_gate_LSTM}
\end{equation}

The cell state can be determined after the input gate activation. Next, the cell state of the previous time step is multiplied element-by-element by the forget gate output. This leads to dismissing values when multiplied by values close to $0$ in the cell state. The input gate output is added element-wise to the cell state. The new cell state in Equation \ref{eq:cell_state_LSTM} is the output:

\begin{equation}
    C(t) = \sigma(f(t)*C(t-1) + i_t)\ .
    \label{eq:cell_state_LSTM}
\end{equation}

The final gate is the output gate, which specifies the next hidden state's value, which includes a certain amount of previous input information. Here the current input and the previous hidden state are summed up and forwarded to the sigmoid function. The new cell state is then transferred to the $\tanh$ function. At the end, the $\tanh$ output with the sigmoid output is multiplied to determine which information the hidden state can carry. The output is a hidden new state. The new cell state and the new hidden state are then shifted to the next stage by Equations \ref{eq:output_gate_LSTM}:

\begin{equation}
    \begin{array}{l}
    o(t) = \sigma(x(t)*U_o + h(t-1)*W_o)\ ,\\
    h(t) = \tanh(C_t)*o(t) \ .
    \end{array}
    \label{eq:output_gate_LSTM}
\end{equation}

To conduct this analysis, we used the Keras framework \cite{chollet2015keras} for deep learning. Our model consists of one stacked LSTM layer and a densely connected output layer with one neuron.

\subsection{Attention Mechanism Neural Network}
\label{subsec:malstm-cfn}

The Attention Function is one of the key aspects of Deep Learning algorithms, an extension of the Encoder-Decoder Paradigm, developed to improve the output on long input sequences.
Figure \ref{fig:AMNN} shows the key idea behind the AMNN, which is to allow the decoder, during decoding, to access encoder information selectively. This is achieved by creating a new context vector for each decoder step, computing it according to the previous hidden state as well as all encoder's hidden states, assigning them trainable weights.
In this way, the Attention Technique gives the input series a different priority and pays more attention to the most important inputs. 

\begin{figure}[htbp]
\centering
\includegraphics[width=0.6\textwidth]{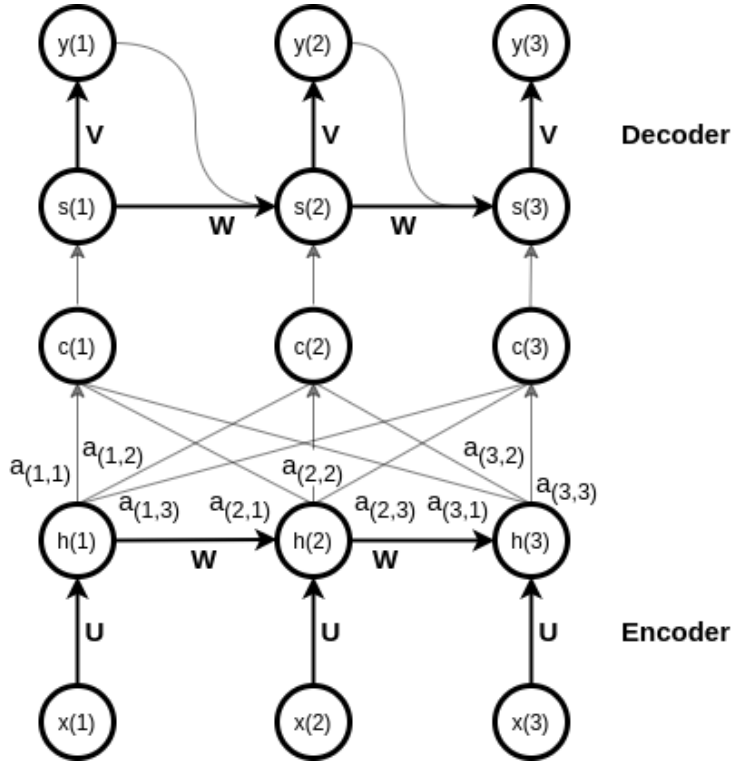}
\caption{Attention Mechanism Neural Network.}
\label{fig:AMNN}
\end{figure}

The encoder operation is very similar to the Encoder-Decoder hybrid operation itself. The representation of each input sequence is determined at each time step, as a function of the previous time step's hidden state and the current input. The final hidden state includes all encoded information from the previous hidden representations and the previous inputs.


The key distinction between the Attention mechanism and the Encoder-Decoder model is that with each decoder step $t$, a new background vector $c(t)$ is computed.
We proceed as follows to measure the context vector $c(t)$ for time step $t$. First of all, the so-called alignment scores $e(j,t)$ are calculated with the weighted sum in Eq. \eqref{eq:context_ANMM} for each combination of the time step $j$ of the encoder and time step $t$ of the decoder:

\begin{equation}
    e(j,t) = V_a*\tanh(U_a*s(t-1) + W_a*h(j))\ .
    \label{eq:context_ANMM}
\end{equation}

$W_a$, $U_a$ and $V_a$ are learning weights in this formula, which are referred to as \textit{attention weights}. The $W_a$ weights are linked to the encoder's hidden states, the $U_a$ weights are linked to the decoder's hidden states, and the $V_a$ weights determine the function that computes the alignment score.
The scores $e(j,t)$ are normalized at each time step $t$ using the softmax function over the time stages of the encoder $j$, obtaining the attention weights $\alpha(j,t)$ as follows:

\begin{equation}
    \alpha(j,t) = \frac{exp(e(j,t))}{\sum_{j=1}^{M}exp(e(j,t))} \ .
    \label{eq:alpha_ANMM}
\end{equation}

 The \textit{importance} of the input at time $j$ is represented by the attention weight $\alpha(j,t)$ for decoding the output of time $t$. The context vector $c(t)$ is estimated according to the attention weights as the weighted sum of all hidden values of the encoder as follows:
\begin{equation}
    c(j,t) = \sum_{j=1}^{M} \alpha(j,t)h(j)\ .
    \label{eq:c_t_ANMM}
\end{equation}

According to this method, the so-called attention function is triggered by the contextual data vector, weighting more the most important inputs.

The contextual vector $c(t)$ is now forwarded to the decoder to calculate the probability distribution for the next possible output. This decoding operation refers to all the time steps present in the input.
The current hidden state $s(t)$ is then calculated according to the recurring unit function, taking as input the contextual vector $c(t)$, the hidden state $s(t-1)$ and the output $\hat{y}(t-1)$ according to the equation:

\begin{equation}
    s(j,t) = f(s(t-1),\hat{y}(t-1),c(t))\ .
    \label{eq:output_ANMM}
\end{equation}

Using this function, the model can identify the relationship between the different parts of the input sequence and the corresponding parts of the output sequence.
The softmax function is used to calculate the output of the decoder in the weighted hidden state at each time $t$:
\begin{equation}
    \hat{y}(t) = softmax(V_s(t))\ .
    \label{eq:output_ANMM}
\end{equation}

Concerning the LSTM, the Attention mechanism provides better results with long input sequences, due to the attention weights. 
 
\begin{figure}[htbp]
\centering
\includegraphics[width=\textwidth]{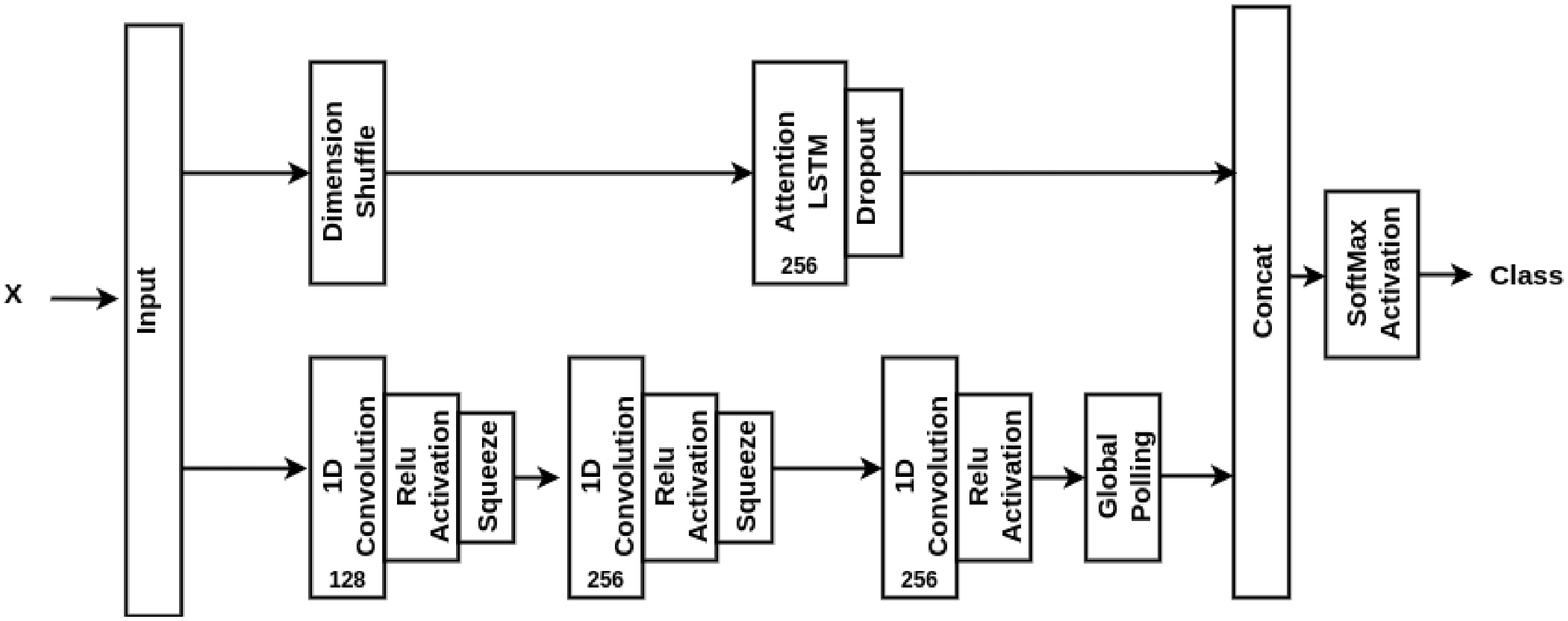}
\caption{Attention LSTM cells to construct the MALSTM-FCN architecture \cite{karim2019multivariate}.}
\label{fig:ALSTM}
\end{figure}

In this study, we specifically use a Multivariate Attention LSTM with Fully Convolutional Network (MALSTM-FCN) proposed by Fazle et al. \cite{karim2019multivariate,karim2017lstm}. Figure \ref{fig:ALSTM} shows the architecture for the MALSTM-FCN including the number of neurons per layer. The input sequence goes in parallel to a fully convolutional layers and Attention LSTM layers, and is concatenated and passed to the output layer via a softmax activation function for binary classification. The fully convolutional block contains three temporal convolutional blocks of 128, 256 and 256 neurons respectively, used as feature extractors. Each convolutional layer is succeeded by batch normalisation, before the concatenation. The dimension shuffle transposes the temporal dimension of the input data, so that the LSTM is given the global temporal information of each variable at once. As a result, the dimension shuffle operation reduces the computation time of training and inference without losing accuracy for time series classification problems \cite{karim2019multivariate}.

\subsection{Convolutional Neural Network}

A Convolutional Neural Network (\textit{CNN}) is a specific class of neural networks most commonly used for deep learning applications concerning image processing, image classification, natural language processing and financial time series analysis \cite{chen2016financial}. 

The most critical part of the CNN architecture is the \textit{convolutional} layer. This layer performs a mathematical operation called \textit{convolution}. In this context, a convolution is a linear operation that involves a multiplication between a matrix of input data and a two-dimensional array of weights, known as a filter. These networks use convolution operation in at least one of their layers.

\begin{figure}[htbp]
\centering
\includegraphics[width=\textwidth]{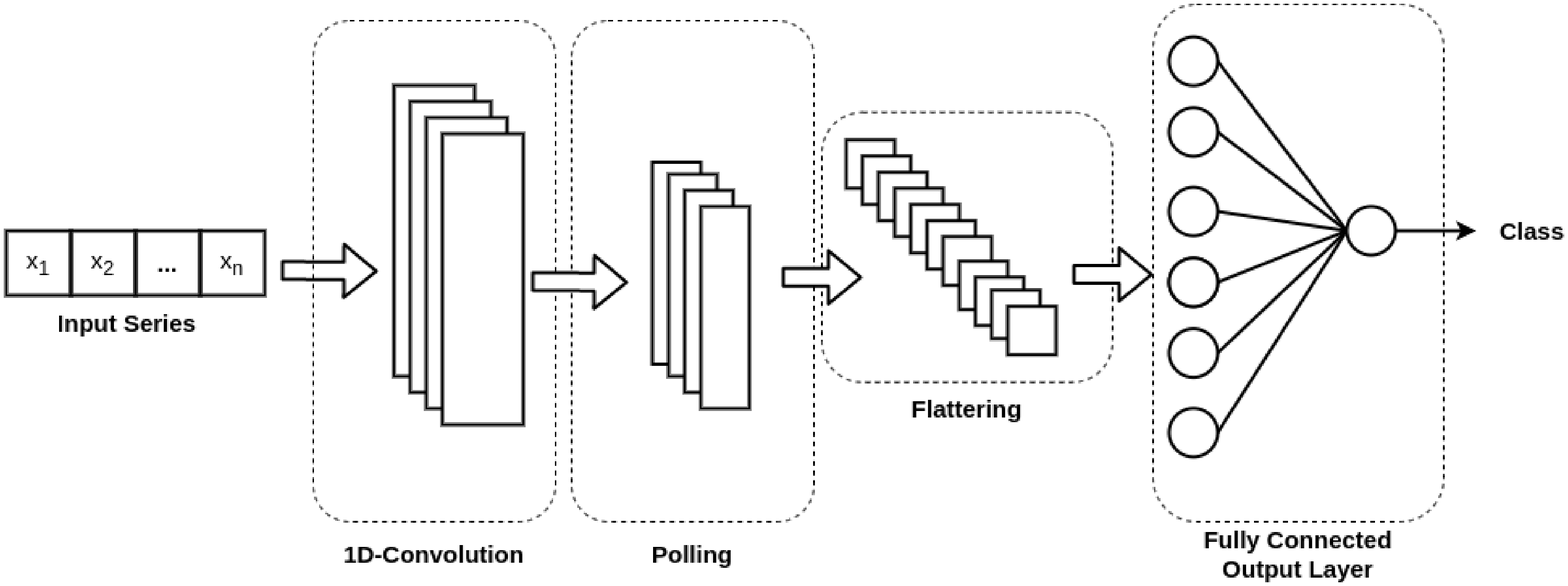}
\caption{Convolutional Neural Network for time series forecasting.}
\label{fig:CNN}
\end{figure}
Convolutional neural networks share a similar architecture with traditional neural networks, including an input and an output layer and multiple hidden layers. The main feature of a CNN is that its hidden layers typically consist of convolutional layers that perform the operations described above. 
Figure \ref{fig:CNN} depicts the general architecture of CNNs for time series analysis. We use a one-dimensional convolutional layer instead of the usual two-dimensional convolutional layer typical in image processing tasks. This first layer is then normalised with a \textit{polling layer} and later \textit{flattened} so that the output layer can process the whole time series at each step $t$. In this case, many one-dimensional convolution layers can be combined in a deep learning network.

For the CNN implementation, we used the Keras framework \cite{chollet2015keras} for deep learning. Our model consists of two or more stacked $1$-dimensional CNN layers, one densely connected layer with $N$ neurons for \textit{polling}, one densely connected layer with $N$ neurons for \textit{flattering}, and finally the densely connected output layer with one neuron.

\subsection{Hyper-parameters tuning}
\label{subsec:hyper_parameters}

The \textit{hyper-parameters tuning} is a method for the optimisation of the hyper-parameters of a given algorithm. It is used to identify the best configuration of the hyper-parameters that would allow the algorithm to achieve the best performance, evaluated with respect to a specific prediction error.
For each algorithm, the hyper-parameters to be optimised are selected, and for each hyper-parameter an appropriate searching interval is defined, including all values to be tested. The algorithm is then fitted on a specific portion of the dataset with the first chosen hyper-parameter configuration. The fitted model is tested on a portion of data that has not been previously used during the training phase. This testing procedure returns a specific value for the chosen prediction error. 
\begin{table}[!htbp]

\begin{center}
\begin{tabular}{cll}
\toprule
\multicolumn{1}{l}{\textbf{Algorithm}} & \textbf{Parameter} & \textbf{Searching Interval}       \\
\toprule
\multirow{6}{*}{MLP}                   & epochs             & 100, 250, 500, 1000               \\
                                       & hidden layers      & 1, 2, 3, 4, 5                     \\
                                       & batch size         & 32, 64, 128, 256, 512             \\
                                       & optimizer          & adam, Nadam, Adamax, RMSprop, SGD \\
                                       & activation         & relu, tanh, softmax               \\
                                       & neurons            & 16, 32, 64, 128, 256              \\
\midrule
\multirow{6}{*}{LSTM}                  & epochs             & 100, 250, 500, 1000               \\
                                       & hidden layers      & 1, 2, 3, 4, 5                     \\
                                       & batch size         & 32, 64, 128, 256, 512             \\
                                       & optimizer          & adam, Nadam, Adamax, RMSprop, SGD \\
                                       & activation         & relu, tanh                        \\
                                       & neurons            & 16, 32, 64, 128, 256              \\
\midrule
\multirow{6}{*}{MALSTM-FCN}            & epochs             & 100, 250, 500, 1000               \\
                                       & hidden layers      & -                                 \\
                                       & batch size         & 32, 64, 128, 256, 512             \\
                                       & optimizer          & adam, Nadam, Adamax, RMSprop, SGD \\
                                       & activation         & -                                 \\
                                       & neurons            & -                                 \\
\midrule
\multirow{6}{*}{CNN}                   & epochs             & 100, 250, 500, 1000               \\
                                       & hidden layers      & 1, 2, 3, 4, 5                     \\
                                       & batch size         & 32, 64, 128, 256, 512             \\
                                       & optimizer          & adam, Nadam, Adamax, RMSprop, SGD \\
                                       & activation         & relu, tanh, softmax               \\
                                       & neurons            & 16, 32, 64, 128, 256          \\ 
\bottomrule
\end{tabular}
\caption{hHyper-parameter searching intervals for different neural network architectures.}
\label{table:tuning}
\end{center}

\end{table}

The optimisation procedure via the Grid Search procedure \cite{lerman1980fitting} ends when all possible combinations of hyper-parameter values have been tested.
The hyper-parameter configuration yielding the best performance in terms of the selected prediction error is therefore chosen as the optimised configuration.
Table \ref{table:tuning} show the hyper-parameters' searching intervals for each implemented algorithm. Since MALSTM-FCN is a deep neural network-specific architecture, the number of layers, neurons per layer and activation function of each layer are already pre-specified (as explained in Section \ref{subsec:malstm-cfn}).

To ensure the robustness of the hyper-parameter optimisation procedure, we use a model validation technique to assess how the performance achieved by a given model will generalise to an independent dataset. This validation technique involves the partition of a data sample into a training set, used to fit the model, and a validation set used to validate the fitted model and a test set to assess the final optimised  generalisation power of the model.
In our analysis, we implemented the \textit{Boostrap Method} \cite{efron1985bootstrap} with 37.8\% of \textit{out-of-bag samples} and 10000 iterations to validate the final hyper-parameters.

\section{Empirical Evidence}
\label{sec:results}

In this section, we report and discuss the main results of the analysis. In particular, we discuss the outcome for both the restricted and unrestricted models. These results are evaluated in terms of the standard classification error metrics: \texttt{accuracy}, \texttt{f1\_score}, \texttt{precision} and \texttt{recall}. 

\subsection{Hyper-Parameters For The Restricted Model}

We briefly discuss here the fine-tuning of the hyper-parameters of the four deep learning algorithm mentioned in Section \ref{subsec:hyper_parameters} considering the hourly frequency resolution. 
Table \ref{table:RestrictedNN} shows the best results obtained for the different neural networks models, using the \textit{Grid Search} technique in terms of the classification error metrics. The best identified parameters with the related results obtained for the \textit{MALSTM-FNC} and \textit{MLP} models are reported in table \ref{table:RestrictedNN}. 

\begin{table}[!htbp]

\begin{center}
\begin{adjustbox}{width={\textwidth},totalheight={\textheight},keepaspectratio}%
\begin{tabular}{cllcccc}
\toprule
\multicolumn{1}{l}{\textbf{Algorithm}} & \textbf{Parameter} & \textbf{Values} & \textbf{Accuracy $(\mu\pm\sigma)$}        & \textbf{Prediction $(\mu\pm\sigma)$}      & \textbf{Recall $(\mu\pm\sigma)$}          & \textbf{f1-score $(\mu\pm\sigma)$}        \\
\toprule
\multirow{6}{*}{MLP}                   & epochs             & 250             & \multirow{6}{*}{0.537 ± 0.029} & \multirow{6}{*}{0.472 ± 0.143} & \multirow{6}{*}{0.511 ± 0.025} & \multirow{6}{*}{0.495 ± 0.027} \\
                                       & hidden layers      & 2               &                                &                                &                                &                                \\
                                       & batch size         & 256,            &                                &                                &                                &                                \\
                                       & optimizer          & Nadam           &                                &                                &                                &                                \\
                                       & activation         & relu            &                                &                                &                                &                                \\
                                       & neurons            & 128             &                                &                                &                                &                                \\
\midrule
\multirow{6}{*}{LSTM}                  & epochs             & 250             & \multirow{6}{*}{0.535 ± 0.034} & \multirow{6}{*}{0.456 ± 0.200} & \multirow{6}{*}{0.485 ± 0.082} & \multirow{6}{*}{0.503 ± 0.285} \\
                                       & hidden layers      & 2               &                                &                                &                                &                                \\
                                       & batch size         & 256,            &                                &                                &                                &                                \\
                                       & optimizer          & Adamx           &                                &                                &                                &                                \\
                                       & activation         & tanh            &                                &                                &                                &                                \\
                                       & neurons            & 256             &                                &                                &                                &                                \\
\midrule
\multirow{6}{*}{MALSTM-FCN}            & epochs             & 250             & \multirow{6}{*}{0.542 ± 0.034} & \multirow{6}{*}{0.456 ± 0.200} & \multirow{6}{*}{0.485 ± 0.082} & \multirow{6}{*}{0.503 ± 0.201} \\
                                       & hidden layers      & -               &                                &                                &                                &                                \\
                                       & batch size         & 256,            &                                &                                &                                &                                \\
                                       & optimizer          & Adamx           &                                &                                &                                &                                \\
                                       & activation         & -               &                                &                                &                                &                                \\
                                       & neurons            & -               &                                &                                &                                &                                \\
\midrule
\multirow{6}{*}{CNN}                   & epochs             & 250             & \multirow{6}{*}{0.435 ± 0.024} & \multirow{6}{*}{0.486 ± 0.210} & \multirow{6}{*}{0.485 ± 0.082} & \multirow{6}{*}{0.453 ± 0.265} \\
                                       & hidden layers      & 2               &                                &                                &                                &                                \\
                                       & batch size         & 128,            &                                &                                &                                &                                \\
                                       & optimizer          & Nadam           &                                &                                &                                &                                \\
                                       & activation         & tanh            &                                &                                &                                &                                \\
                                       & neurons            & 128             &                                &                                &                                &     \\
\bottomrule
\end{tabular}
\end{adjustbox}
\end{center}
\caption{Restricted model - Neural networks optimal parameters.}
\label{table:RestrictedNN}
\end{table}

The neural network that achieved the best accuracy is \textit{MALSTM-FNC}, with an average accuracy of $53.7\%$ and a standard deviation of $2.9\%$. Among the implemented machine learning models, the one that achieved the best f1-score is again \textit{MALSTM-FNC}, with an average accuracy of $54\%$ and a standard deviation of $2.01\%$ (the \textit{LSTM} obtained the same f1-score but we observe a higher variance).

\subsection{Hyper-Parameters For The Unrestricted Model}

Table \ref{table:UnrestrictedNN} shows the best results obtained for the Neural Networks models, via the \textit{Grid Search} technique with respect to the classification error metrics. The best identified parameters with the related results obtained for the \textit{CNN} and \textit{LSTM} models are reported in table \ref{table:UnrestrictedNN}.

\begin{table}[!htbp]

\begin{center}
\begin{adjustbox}{width={\textwidth},totalheight={\textheight},keepaspectratio}%
\begin{tabular}{cllcccc}
\toprule
\multicolumn{1}{l}{\textbf{Algorithm}} & \textbf{Parameter} & \textbf{Values} & \textbf{Accuracy $(\mu\pm\sigma)$} & \textbf{Prediction $(\mu\pm\sigma)$} & \textbf{Recall $(\mu\pm\sigma)$} & \textbf{f1-score $(\mu\pm\sigma)$} \\
\toprule
\multirow{6}{*}{MLP}                   & epochs             & 500             & \multirow{6}{*}{0.81 ± 0.025}      & \multirow{6}{*}{0.984 ± 0.180}       & \multirow{6}{*}{0.541 ± 0.060}   & \multirow{6}{*}{0.698 ± 0.908}     \\
                                       & hidden layers      & 3               &                                    &                                      &                                  &                                    \\
                                       & batch size         & 256             &                                    &                                      &                                  &                                    \\
                                       & optimizer          & Nadam           &                                    &                                      &                                  &                                    \\
                                       & activation         & relu            &                                    &                                      &                                  &                                    \\
                                       & neurons            & 128             &                                    &                                      &                                  &                                    \\
\midrule
\multirow{6}{*}{LSTM}                  & epochs             & 1000             & \multirow{6}{*}{0.86 ± 0.027}      & \multirow{6}{*}{0.918 ± 0.033}       & \multirow{6}{*}{0.873 ± 0.228}   & \multirow{6}{*}{0.895 ± 0.175}     \\
                                       & hidden layers      & 3               &                                    &                                      &                                  &                                    \\
                                       & batch size         & 256             &                                    &                                      &                                  &                                    \\
                                       & optimizer          & Adamx           &                                    &                                      &                                  &                                    \\
                                       & activation         & tanh            &                                    &                                      &                                  &                                    \\
                                       & neurons            & 256             &                                    &                                      &                                  &                                    \\
\midrule
\multirow{6}{*}{MALSTM-FCN}            & epochs             & 500             & \multirow{6}{*}{0.73 ± 0.027}      & \multirow{6}{*}{0.75 ± 0.241}        & \multirow{6}{*}{0.611 ± 0.175}   & \multirow{6}{*}{0.673 ± 0.060}     \\
                                       & hidden layers      & -               &                                    &                                      &                                  &                                    \\
                                       & batch size         & 256             &                                    &                                      &                                  &                                    \\
                                       & optimizer          & Adamx           &                                    &                                      &                                  &                                    \\
                                       & activation         & -               &                                    &                                      &                                  &                                    \\
                                       & neurons            & -               &                                    &                                      &                                  &                                    \\
\midrule
\multirow{6}{*}{CNN}                   & epochs             & 1000            & \multirow{6}{*}{0.87 ± 0.027}      & \multirow{6}{*}{0.782 ± 0.175}       & \multirow{6}{*}{0.913 ± 0.060}   & \multirow{6}{*}{0.842 ± 0.228}     \\
                                       & hidden layers      & 2               &                                    &                                      &                                  &                                    \\
                                       & batch size         & 256             &                                    &                                      &                                  &                                    \\
                                       & optimizer          & Nadam           &                                    &                                      &                                  &                                    \\
                                       & activation         & relu            &                                    &                                      &                                  &                                    \\
                                       & neurons            & 256             &                                    &                                      &                                  &                                   \\
\bottomrule
\end{tabular}
\end{adjustbox}
\end{center}
\caption{Unrestricted model - Neural networks optimal parameters.}
\label{table:UnrestrictedNN}
\end{table}

The results obtained for the unrestricted model highlight that the addition of trading and social media indicators to the model leads to an effective improvement in average accuracy, namely the prediction error.
This result is consistent for all implemented algorithms, and this allows us to exclude that this result is a statistical fluctuation or that it may be an artefact of the particular classification algorithm implemented.
The best result obtained with the unrestricted model is achieved using the \textit{CNN} model, with a mean accuracy of $87\%$ and a standard deviation of $2.7\%$.

\subsection{Results and Discussions}
\label{sebsec:result_and_discuss}

Table \ref{tab:resutls_all_hourly} shows the results obtained using the four deep learning algorithms for the \textbf{hourly} frequency price movements classification task. This table presents the results for the restricted (upper part) and unrestricted (lower part) model. Firstly, it can be noted that for all four deep learning algorithms, the performance of the unrestricted model outperforms the restricted model in terms of accuracy, precision, recall and F1 score. The accuracy ranges from 51\% of for the restricted MLP to 84\%  for CNNs and LSTMs.

The fact that the result is consistent across
all four classifiers, further confirm that is not due to statistical fluctuations, but rather to the higher predictive of the unrestricted model. For Bitcoin, the highest performances are obtained using the CNN architecture and for Ethereum by the LSTM.

We have also further explored the classification via the unrestricted model at hourly frequency considering two sub-models: a sub-model including technical and social indicators and the other with all the indicators (social, technical and trading). In this way, it is possible to disentangle impact of social and trading indicators on the models' performance. We used a statistical $t$ test on the distributions of accuracy, prediction, recall and F1-score for the two unrestricted sub-modules finding that adding social indicators does not add a significant improvement to the unrestricted model. For this reason, in Table \ref{tab:resutls_all_hourly} we omitted the unrestricted model including social and technical indicators only.

\begin{table}[!htp]
\centering
\begin{adjustbox}{max width=\textwidth}
\begin{tabular}{c|c|c|l|cccc}
\toprule
\textbf{Model} & \textbf{Algotithm} & \textbf{Cryptocurrency} & \textbf{Class}   & \multicolumn{1}{c}{\textbf{Accuracy}} & \multicolumn{1}{c}{\textbf{Precision}} & \multicolumn{1}{c}{\textbf{Recall}} & \multicolumn{1}{c}{\textbf{F1-score}} \\
\toprule
\multirow{24}{*}{\textbf{Restricted}}   & \multirow{6}{*}{\textbf{MLP}}   & \multirow{3}{*}{\textbf{bitcoin}}  & \textbf{down} & & 0.57 & 0.28 & 0.38 \\
    & & & \textbf{up} & & 0.54 & 0.80 & 0.64 \\
    & & & \textbf{average} & \multicolumn{1}{c}{\cellcolor[HTML]{E06666}0.55}       & \cellcolor[HTML]{E06666}0.56 & \cellcolor[HTML]{E06666}0.55 
    & \cellcolor[HTML]{E06666}0.51 \\
    &  & \multirow{3}{*}{\textbf{ethereum}} & \textbf{down}    &                    & 0.52 & 0.77 & 0.62 \\
                                        &                                 &                                    & \textbf{up}      &                                       & 0.55                                   & 0.29                                & 0.38                                  \\
                                        &                                 &                                    & \textbf{average} & \multicolumn{1}{c}{\cellcolor[HTML]{E06666}0.53}              & \cellcolor[HTML]{E06666}0.54                                   & \cellcolor[HTML]{E06666}0.53                                & \cellcolor[HTML]{E06666}0.50                                  \\
\cline{2-8}
                                        & \multirow{6}{*}{\textbf{MALSTM-FNC}} & \multirow{3}{*}{\textbf{bitcoin}}  & \textbf{down}    &                                       & 0.52                                   & 0.50                                & 0.51                                  \\
                                        &                                 &                                    & \textbf{up}      &                                       & 0.55                                   & 0.57                                & 0.56                                  \\
                                        &                                 &                                    & \textbf{average} & \multicolumn{1}{c}{\cellcolor[HTML]{E06666}0.54}              & \cellcolor[HTML]{E06666}0.54                                   & \cellcolor[HTML]{E06666}0.54                                & \cellcolor[HTML]{E06666}0.54                                  \\
                                        &                                 & \multirow{3}{*}{\textbf{ethereum}} & \textbf{down}    &                                       & 0.52                                   & 0.80                                & 0.63                                  \\
                                        &                                 &                                    & \textbf{up}      &                                       & 0.57                                   & 0.26                                & 0.36                                  \\
                                        &                                 &                                    & \textbf{average} & \multicolumn{1}{c}{\cellcolor[HTML]{E06666}0.53}              & \cellcolor[HTML]{E06666}0.54                                   & \cellcolor[HTML]{E06666}0.53                                & \cellcolor[HTML]{E06666}0.50                                  \\
\cline{2-8}
                                        & \multirow{6}{*}{\textbf{LSTM}}  & \multirow{3}{*}{\textbf{bitcoin}}  & \textbf{down}    &                                       & 0.49                                   & 0.29                                & 0.37                                  \\
                                        &                                 &                                    & \textbf{up}      &                                       & 0.53                                   & 0.73                                & 0.61                                  \\
                                        &                                 &                                    & \textbf{average} & \multicolumn{1}{c}{\cellcolor[HTML]{E06666}0.52}              & \cellcolor[HTML]{E06666}0.51                                   & \cellcolor[HTML]{E06666}0.52                                & \cellcolor[HTML]{E06666}0.49                                  \\
                                        &                                 & \multirow{3}{*}{\textbf{ethereum}} & \textbf{down}    &                                       & 0.51                                   & 0.70                                & 0.59                                  \\
                                        &                                 &                                    & \textbf{up}      &                                       & 0.51                                   & 0.31                                & 0.39                                  \\
                                        &                                 &                                    & \textbf{average} & \multicolumn{1}{c}{\cellcolor[HTML]{E06666}0.51}              & \cellcolor[HTML]{E06666}0.51                                   & \cellcolor[HTML]{E06666}0.51                                & \cellcolor[HTML]{E06666}0.49                                  \\
\cline{2-8}
                                        & \multirow{6}{*}{\textbf{CNN}}   & \multirow{3}{*}{\textbf{bitcoin}}  & \textbf{down}    &                                       & 0.52                                   & 0.65                                & 0.57                                  \\
                                        &                                 &                                    & \textbf{up}      &                                       & 0.56                                   & 0.42                                & 0.48                                  \\
                                        &                                 &                                    & \textbf{average} & \multicolumn{1}{c}{\cellcolor[HTML]{E06666}0.53}              & \cellcolor[HTML]{E06666}0.54                                   & \cellcolor[HTML]{E06666}0.53                                & \cellcolor[HTML]{E06666}0.53                                  \\
                                        &                                 & \multirow{3}{*}{\textbf{ethereum}} & \textbf{down}    &                                       & 0.50                                   & 0.75                                & 0.60                                  \\
                                        &                                 &                                    & \textbf{up}      &                                       & 0.56                                   & 0.31                                & 0.40                                  \\
                                        &                                 &                                    & \textbf{average} & \multicolumn{1}{c}{\cellcolor[HTML]{E06666}0.52}              & \cellcolor[HTML]{E06666}0.53                                   & \cellcolor[HTML]{E06666}0.52                                &\cellcolor[HTML]{E06666} 0.49                                  \\
\midrule
\multirow{24}{*}{\textbf{Unrestricted}} & \multirow{6}{*}{\textbf{MLP}}   & \multirow{3}{*}{\textbf{bitcoin}}  & \textbf{down}    &                                       & 0.87                                   & 0.57                                & 0.69                                  \\
                                        &                                 &                                    & \textbf{up}      &                                       & 0.70                                   & 0.92                                & 0.79                                  \\
                                        &                                 &                                    & \textbf{average} & \multicolumn{1}{c}{\textbf{\cellcolor[HTML]{93C47D}0.75}}              & \cellcolor[HTML]{93C47D}0.78                                   & \cellcolor[HTML]{93C47D}0.75                                & \cellcolor[HTML]{93C47D}0.74                                  \\
                                        &                                 & \multirow{3}{*}{\textbf{ethereum}} & \textbf{down}    &                                       & 0.80                                   & 0.79                                & 0.80                                  \\
                                        &                                 &                                    & \textbf{up}      &                                       & 0.80                                   & 0.80                                & 0.80                                  \\
                                        &                                 &                                    & \textbf{average} & \multicolumn{1}{c}{\cellcolor[HTML]{93C47D}\textbf{0.80}}              & \cellcolor[HTML]{93C47D}0.80                                   & \cellcolor[HTML]{93C47D}0.80                                & \cellcolor[HTML]{93C47D}0.80                                  \\
\cline{2-8}
                                        & \multirow{6}{*}{\textbf{MALSTM-FNC}} & \multirow{3}{*}{\textbf{bitcoin}}  & \textbf{down}    &                                       & 0.97                                   & 0.32                                & 0.48                                  \\
                                        &                                 &                                    & \textbf{up}      &                                       & 0.61                                   & 0.99                                & 0.75                                  \\
                                        &                                 &                                    & \textbf{average} & \multicolumn{1}{c}{\cellcolor[HTML]{93C47D}\textbf{0.67}}              & \cellcolor[HTML]{93C47D}0.78                                   & \cellcolor[HTML]{93C47D}0.67                                & \cellcolor[HTML]{93C47D}0.62                                  \\
                                        &                                 & \multirow{3}{*}{\textbf{ethereum}} & \textbf{down}    &                                       & 0.98                                   & 0.15                                & 0.27                                  \\
                                        &                                 &                                    & \textbf{up}      &                                       & 0.54                                   & 1.00                                & 0.70                                  \\
                                        &                                 &                                    & \textbf{average} & \multicolumn{1}{c}{\cellcolor[HTML]{93C47D}\textbf{0.58}}              & \cellcolor[HTML]{93C47D}0.76                                   & \cellcolor[HTML]{93C47D}0.58                                & \cellcolor[HTML]{93C47D}0.49                                  \\
\cline{2-8}
                                        & \multirow{6}{*}{\textbf{LSTM}}  & \multirow{3}{*}{\textbf{bitcoin}}  & \textbf{down}    &                                       & 0.79                                   & 0.90                                & 0.84                                  \\
                                        &                                 &                                    & \textbf{up}      &                                       & 0.88                                   & 0.76                                & 0.82                                  \\
                                        &                                 &                                    & \textbf{average} & \multicolumn{1}{c}{\cellcolor[HTML]{93C47D}\textbf{0.83}}              & \cellcolor[HTML]{93C47D}0.84                                   & \cellcolor[HTML]{93C47D}0.83                                & \cellcolor[HTML]{93C47D}0.83                                  \\
                                        &                                 & \multirow{3}{*}{\textbf{ethereum}} & \textbf{down}    &                                       & 0.79                                   & 0.91                                & 0.84                                  \\
                                        &                                 &                                    & \textbf{up}      &                                       & 0.90                                   & 0.76                                & 0.83                                  \\
                                        &                                 &                                    & \textbf{average} & \multicolumn{1}{c}{\cellcolor[HTML]{93C47D}\textbf{0.84}}              & \cellcolor[HTML]{93C47D}0.84                                   & \cellcolor[HTML]{93C47D}0.84                                & \cellcolor[HTML]{93C47D}0.83                                  \\
\cline{2-8}
                                        & \multirow{6}{*}{\textbf{CNN}}   & \multirow{3}{*}{\textbf{bitcoin}}  & \textbf{down}    &                                       & 0.82                                   & 0.87                                & 0.84                                  \\
                                        &                                 &                                    & \textbf{up}      &                                       & 0.87                                   & 0.82                                & 0.85                                  \\
                                        &                                 &                                    & \textbf{average} & \multicolumn{1}{c}{\cellcolor[HTML]{93C47D}\textbf{0.84}}              & \cellcolor[HTML]{93C47D}0.84                                   & \cellcolor[HTML]{93C47D}0.84                                & \cellcolor[HTML]{93C47D}0.84                                  \\
                                        &                                 & \multirow{3}{*}{\textbf{ethereum}} & \textbf{down}    &                                       & 0.72                                   & 0.97                                & 0.83                                  \\
                                        &                                 &                                    & \textbf{up}      &                                       & 0.95                                   & 0.61                                & 0.74                                  \\
                                        &                                 &                                    & \textbf{average} & \multicolumn{1}{c}{\cellcolor[HTML]{93C47D}\textbf{0.79}}             
                                        & \cellcolor[HTML]{93C47D}0.83                                   & \cellcolor[HTML]{93C47D}0.79 
                                        & \cellcolor[HTML]{93C47D}0.78\\

\bottomrule
\end{tabular}
\end{adjustbox}
\caption{Accuracy, Precision, Recall, F1 score for Restricted and Unrestricted models for each Deep Learning Algorithm With Hourly Frequency.}
\label{tab:resutls_all_hourly}
\end{table}

\begin{table}[]
\begin{adjustbox}{max width=0.9\textwidth}
\begin{tabular}{c|p{2cm}|c|c|r|lrrr}
\toprule
\textbf{Model}                           & \textbf{Features}                                        & \textbf{Algotithm}               & \textbf{Cryptocurrency}             & \multicolumn{1}{c}{\textbf{Class}} & \multicolumn{1}{c}{\textbf{Accuracy}}                     & \multicolumn{1}{c}{\textbf{Precision}} & \multicolumn{1}{c}{\textbf{Recall}}   & \multicolumn{1}{c}{\textbf{F1-score}} \\
\toprule
                                         &                                                          &                                  &                                     & \textbf{down}                      &                                                           & 0.00                                   & 0.00                                  & 0.00                                  \\
                                         &                                                          &                                  &                                     & \textbf{up}                        &                                                           & 0.59                                   & 0.96                                  & 0.73                                  \\
                                         &                                                          &                                  & \multirow{-3}{*}{\textbf{bitcoin}}  & \textbf{average}                   & \multicolumn{1}{r}{\textbf{0.58}}                         & \textbf{0.36}                          & \textbf{0.58}                         & \textbf{0.44}                         \\
                                         &                                                          &                                  &                                     & \textbf{down}                      &                                                           & 0.96                                   & 1.00                                  & 0.98                                  \\
                                         &                                                          &                                  &                                     & \textbf{up}                        &                                                           & 1.00                                   & 0.96                                  & 0.98                                  \\
                                         &                                                          & \multirow{-6}{*}{\textbf{MLP}}   & \multirow{-3}{*}{\textbf{ethereum}} & \textbf{average}                   & \multicolumn{1}{r}{\textbf{0.98}}                         & \textbf{0.98}                          & \textbf{0.98}                         & \textbf{0.98}                         \\
\cline{3-9}
                                         &                                                          &                                  &                                     & \textbf{down}                      &                                                           & 0.51                                   & 0.47                                  & 0.49                                  \\
                                         &                                                          &                                  &                                     & \textbf{up}                        &                                                           & 0.56                                   & 0.59                                  & 0.58                                  \\
                                         &                                                          &                                  & \multirow{-3}{*}{\textbf{bitcoin}}  & \textbf{average}                   & \multicolumn{1}{r}{\textbf{0.54}}                         & \textbf{0.54}                          & \textbf{0.54}                         & \textbf{0.54}                         \\
                                         &                                                          &                                  &                                     & \textbf{down}                      &                                                           & 1.00                                   & 0.99                                  & 0.99                                  \\
                                         &                                                          &                                  &                                     & \textbf{up}                        &                                                           & 0.99                                   & 1.00                                  & 0.99                                  \\
                                         &                                                          & \multirow{-6}{*}{\textbf{MALSTM-FNC}} & \multirow{-3}{*}{\textbf{ethereum}} & \textbf{average}                   & \multicolumn{1}{r}{\cellcolor[HTML]{93C47D}\textbf{0.99}} & \cellcolor[HTML]{93C47D}\textbf{0.99}  & \cellcolor[HTML]{93C47D}\textbf{0.99} & \cellcolor[HTML]{93C47D}\textbf{0.99} \\
\cline{3-9}
                                         &                                                          &                                  &                                     & \textbf{down}                      &                                                           & 0.00                                   & 0.00                                  & 0.00                                  \\
                                         &                                                          &                                  &                                     & \textbf{up}                        &                                                           & 0.57                                   & 1.00                                  & 0.73                                  \\
                                         &                                                          &                                  & \multirow{-3}{*}{\textbf{bitcoin}}  & \textbf{average}                   & \multicolumn{1}{r}{\textbf{0.57}}                         & \textbf{0.33}                          & \textbf{0.57}                         & \textbf{0.41}                         \\
                                         &                                                          &                                  &                                     & \textbf{down}                      &                                                           & 0.98                                   & 0.98                                  & 0.98                                  \\
                                         &                                                          &                                  &                                     & \textbf{up}                        &                                                           & 0.99                                   & 0.99                                  & 0.99                                  \\
                                         &                                                          & \multirow{-6}{*}{\textbf{LSTM}}  & \multirow{-3}{*}{\textbf{ethereum}} & \textbf{average}                   & \multicolumn{1}{r}{\textbf{0.99}}                         & \textbf{0.99}                          & \textbf{0.99}                         & \textbf{0.99}                         \\
\cline{3-9}
                                         &                                                          &                                  &                                     & \textbf{down}                      &                                                           & 0.38                                   & 0.10                                  & 0.16                                  \\
                                         &                                                          &                                  &                                     & \textbf{up}                        &                                                           & 0.60                                   & 0.89                                  & 0.72                                  \\
                                         &                                                          &                                  & \multirow{-3}{*}{\textbf{bitcoin}}  & \textbf{average}                   & \multicolumn{1}{r}{\textbf{0.58}}                         & \textbf{0.51}                          & \textbf{0.58}                         & \textbf{0.50}                         \\
                                         &                                                          &                                  &                                     & \textbf{down}                      &                                                           & 0.88                                   & 1.00                                  & 0.94                                  \\
                                         &                                                          &                                  &                                     & \textbf{up}                        &                                                           & 1.00                                   & 0.88                                  & 0.94                                  \\

\multirow{-24}{*}{\textbf{Restricted}}   & \multirow{-24}{*}{\textbf{technical}}                    & \multirow{-6}{*}{\textbf{CNN}}   & \multirow{-3}{*}{\textbf{ethereum}} & \textbf{average}                   & \multicolumn{1}{r}{\textbf{0.94}}                         & \textbf{0.94}                          & \textbf{0.94}                         & \textbf{0.94}                         \\
\midrule
                                         &                                                          &                                  &                                     & \textbf{down}                      &                                                           & 0.59                                   & 0.21                                  & 0.31                                  \\
                                         &                                                          &                                  &                                     & \textbf{up}                        &                                                           & 0.61                                   & 0.90                                  & 0.72                                  \\
                                         &                                                          &                                  & \multirow{-3}{*}{\textbf{bitcoin}}  & \textbf{average}                   & \multicolumn{1}{r}{\cellcolor[HTML]{93C47D}\textbf{0.60}} & \cellcolor[HTML]{93C47D}\textbf{0.60}  & \cellcolor[HTML]{93C47D}\textbf{0.60} & \cellcolor[HTML]{93C47D}\textbf{0.55} \\
                                         &                                                          &                                  &                                     & \textbf{down}                      &                                                           & 0.79                                   & 0.95                                  & 0.87                                  \\
                                         &                                                          &                                  &                                     & \textbf{up}                        &                                                           & 0.95                                   & 0.79                                  & 0.87                                  \\
                                         &                                                          & \multirow{-6}{*}{\textbf{MLP}}   & \multirow{-3}{*}{\textbf{ethereum}} & \textbf{average}                   & \multicolumn{1}{r}{0.87}                                  & 0.88                                   & 0.87                                  & 0.87                                  \\
\cline{3-9}
                                         &                                                          &                                  &                                     & \textbf{down}                      &                                                           & 0.41                                   & 0.41                                  & 0.41                                  \\
                                         &                                                          &                                  &                                     & \textbf{up}                        &                                                           & 0.51                                   & 0.51                                  & 0.51                                  \\
                                         &                                                          &                                  & \multirow{-3}{*}{\textbf{bitcoin}}  & \textbf{average}                   & \multicolumn{1}{r}{\textbf{0.46}}                         & \textbf{0.46}                          & \textbf{0.46}                         & \textbf{0.46}                         \\
                                         &                                                          &                                  &                                     & \textbf{down}                      &                                                           & 0.72                                   & 0.70                                  & 0.71                                  \\
                                         &                                                          &                                  &                                     & \textbf{up}                        &                                                           & 0.77                                   & 0.78                                  & 0.77                                  \\
                                         &                                                          & \multirow{-6}{*}{\textbf{MALSTM-FNC}} & \multirow{-3}{*}{\textbf{ethereum}} & \textbf{average}                   & \multicolumn{1}{r}{\textbf{0.75}}                         & \textbf{0.75}                          & \textbf{0.75}                         & \textbf{0.75}                         \\
\cline{3-9}
                                         &                                                          &                                  &                                     & \textbf{down}                      &                                                           & 0.44                                   & 0.10                                  & 0.17                                  \\
                                         &                                                          &                                  &                                     & \textbf{up}                        &                                                           & 0.47                                   & 0.86                                  & 0.60                                  \\
                                         &                                                          &                                  & \multirow{-3}{*}{\textbf{bitcoin}}  & \textbf{average}                   & \multicolumn{1}{r}{\textbf{0.46}}                         & \textbf{0.45}                          & \textbf{0.46}                         & \textbf{0.38}                         \\
                                         &                                                          &                                  &                                     & \textbf{down}                      &                                                           & 0.88                                   & 0.83                                  & 0.85                                  \\
                                         &                                                          &                                  &                                     & \textbf{up}                        &                                                           & 0.87                                   & 0.91                                  & 0.89                                  \\
                                         &                                                          & \multirow{-6}{*}{\textbf{LSTM}}  & \multirow{-3}{*}{\textbf{ethereum}} & \textbf{average}                   & \multicolumn{1}{r}{\textbf{0.87}}                         & \textbf{0.87}                          & \textbf{0.87}                         & \textbf{0.87}                         \\
\cline{3-9}
                                         &                                                          &                                  &                                     & \textbf{down}                      &                                                           & 0.42                                   & 0.47                                  & 0.44                                  \\
                                         &                                                          &                                  &                                     & \textbf{up}                        &                                                           & 0.56                                   & 0.52                                  & 0.54                                  \\
                                         &                                                          &                                  & \multirow{-3}{*}{\textbf{bitcoin}}  & \textbf{average}                   & \multicolumn{1}{r}{\textbf{0.50}}                         & \textbf{0.50}                          & \textbf{0.50}                         & \textbf{0.50}                         \\
                                         &                                                          &                                  &                                     & \textbf{down}                      &                                                           & 0.77                                   & 0.83                                  & 0.80                                  \\
                                         &                                                          &                                  &                                     & \textbf{up}                        &                                                           & 0.85                                   & 0.79                                  & 0.82                                  \\
                                         & \multirow{-24}{2cm}{\textbf{technical + social}}           & \multirow{-6}{*}{\textbf{CNN}}   & \multirow{-3}{*}{\textbf{ethereum}} & \textbf{average}                   & \multicolumn{1}{r}{\textbf{0.81}}                         & \textbf{0.81}                          & \textbf{0.81}                         & \textbf{0.81}                         \\
\cline{2-9}
                                         &                                                          &                                  &                                     & \textbf{down}                      &                                                           & 0.59                                   & 0.20                                  & 0.30                                  \\
                                         &                                                          &                                  &                                     & \textbf{up}                        &                                                           & 0.47                                   & 0.84                                  & 0.60                                  \\
                                         &                                                          &                                  & \multirow{-3}{*}{\textbf{bitcoin}}  & \textbf{average}                   & \multicolumn{1}{r}{\textbf{0.49}}                         & \textbf{0.54}                          & \textbf{0.49}                         & \textbf{0.43}                         \\
                                         &                                                          &                                  &                                     & \textbf{down}                      &                                                           & 0.84                                   & 0.91                                  & 0.87                                  \\
                                         &                                                          &                                  &                                     & \textbf{up}                        &                                                           & 0.91                                   & 0.84                                  & 0.87                                  \\
                                         &                                                          & \multirow{-6}{*}{\textbf{MLP}}   & \multirow{-3}{*}{\textbf{ethereum}} & \textbf{average}                   & \multicolumn{1}{r}{\textbf{0.87}}                         & \textbf{0.88}                          & \textbf{0.87}                         & \textbf{0.87}                         \\
\cline{3-9}
                                         &                                                          &                                  &                                     & \textbf{down}                      &                                                           & 0.41                                   & 0.41                                  & 0.41                                  \\
                                         &                                                          &                                  &                                     & \textbf{up}                        &                                                           & 0.62                                   & 0.62                                  & 0.62                                  \\
                                         &                                                          &                                  & \multirow{-3}{*}{\textbf{bitcoin}}  & \textbf{average}                   & \multicolumn{1}{r}{\textbf{0.54}}                         & \textbf{0.54}                          & \textbf{0.54}                         & \textbf{0.54}                         \\
                                         &                                                          &                                  &                                     & \textbf{down}                      &                                                           & 0.79                                   & 0.88                                  & 0.83                                  \\
                                         &                                                          &                                  &                                     & \textbf{up}                        &                                                           & 0.91                                   & 0.83                                  & 0.87                                  \\
                                         &                                                          & \multirow{-6}{*}{\textbf{MALSTM-FNC}} & \multirow{-3}{*}{\textbf{ethereum}} & \textbf{average}                   & \multicolumn{1}{r}{\textbf{0.85}}                         & \textbf{0.86}                          & \textbf{0.85}                         & \textbf{0.85}                         \\
\cline{3-9}
                                         &                                                          &                                  &                                     & \textbf{down}                      &                                                           & 0.44                                   & 0.31                                  & 0.36                                  \\
                                         &                                                          &                                  &                                     & \textbf{up}                        &                                                           & 0.43                                   & 0.58                                  & 0.49                                  \\
                                         &                                                          &                                  & \multirow{-3}{*}{\textbf{bitcoin}}  & \textbf{average}                   & \multicolumn{1}{r}{\textbf{0.44}}                         & \textbf{0.44}                          & \textbf{0.44}                         & \textbf{0.43}                         \\
                                         &                                                          &                                  &                                     & \textbf{down}                      &                                                           & 0.92                                   & 0.87                                  & 0.89                                  \\
                                         &                                                          &                                  &                                     & \textbf{up}                        &                                                           & 0.86                                   & 0.91                                  & 0.88                                  \\
                                         &                                                          & \multirow{-6}{*}{\textbf{LSTM}}  & \multirow{-3}{*}{\textbf{ethereum}} & \textbf{average}                   & \multicolumn{1}{r}{\textbf{0.89}}                         & \textbf{0.89}                          & \textbf{0.89}                         & \textbf{0.89}                         \\
\cline{3-9}
                                         &                                                          &                                  &                                     & \textbf{down}                      &                                                           & 0.52                                   & 0.55                                  & 0.54                                  \\
                                         &                                                          &                                  &                                     & \textbf{up}                        &                                                           & 0.62                                   & 0.59                                  & 0.60                                  \\
                                         &                                                          &                                  & \multirow{-3}{*}{\textbf{bitcoin}}  & \textbf{average}                   & \multicolumn{1}{r}{\textbf{0.57}}                         & \textbf{0.57}                          & \textbf{0.57}                         & \textbf{0.57}                         \\
                                         &                                                          &                                  &                                     & \textbf{down}                      &                                                           & 0.92                                   & 0.85                                  & 0.89                                  \\
                                         &                                                          &                                  &                                     & \textbf{up}                        &                                                           & 0.87                                   & 0.93                                  & 0.90                                  \\
\multirow{-48}{*}{\textbf{Unrestricted}} & \multirow{-24}{2cm}{\textbf{technical + social + trading}} & \multirow{-6}{*}{\textbf{CNN}}   & \multirow{-3}{*}{\textbf{ethereum}} & \textbf{average}                   & \multicolumn{1}{r}{\textbf{0.89}}                         & \textbf{0.90}                          & \textbf{0.89}                         & \textbf{0.89}      \\
\bottomrule
\end{tabular}
\end{adjustbox}
\caption{Accuracy, Precision, Recall, F1 score for Restricted and Unrestricted models for each Deep Learning Algorithm For Daily Frequency.}
\label{tab:resutls_all_daily}
\end{table}

Table \ref{tab:resutls_all_daily} shows the results obtained by the four deep learning algorithms for \textbf{daily} frequency price movements classification. This table presents results for both the restricted (upper part) and unrestricted (lower part) model. The unrestricted model is further divided in \textit{technical-social} and \textit{techical-social-trading} sub-models to better highlight the contribution of social and trading indicators to the model separately.

The MALSTM-CNF achieves the best classification performance for Ethereum with 99\% of accuracy using the restricted model composed of only technical indicators. For Bitcoin, the best results are achieved by MLP with F1-score of 55\% and accuracy of 60\% with the unrestricted model with \textbf{only social media indicators} and technical indicators (in this case, we consider F1-score and accuracy for Bitcoin because of the slightly unbalanced class distribution described in Section \ref{sub:price_movement_classification}).
For the daily frequency classification, we can see that in general technical indicators alone performs better in the classification of next day price movement. The more indicators we add to the model, the more the performance decrease. 
Another general result is that the accuracy, precision, recall and F1-score for daily classification of Ethereum price movements are far better than those for Bitcoin.
Results for daily classification are in line with other studies \cite{akyildirim2020prediction} for the hourly and daily classification with a significant improvement when considering the hourly unrestricted model. The social media indicators turn out to be particularly relevant at the daily frequency for the Bitcoin case. This result is in agreement with the recent result on the impact social media sentiment on cryptocurrencies markets \cite{bartolucci2020butterfly}: the effects of social media on markets show a long lag, which is not captured nor relevant at an hourly frequency.

\section{Threats To Validity} 
\label{sec:threats}
In this section, we discuss potential limitations and threats to validity of our analysis. First, our analysis focuses on Ethereum and Bitcoin: this may constitute a threat to external validity as conducting the analysis for different cryptocurrencies may lead to different results.

Secondly, threats to internal validity concern confounding factors that can influence the obtained results. Based on the empirical evidence, we assume that technical, trading and social indicators are exhaustive in the case of our model. There may exists nonetheless other factors omitted from  this study, which could influence the price movements. 

Finally, threats to construct validity focus on how accurately the observations describe the phenomena of interest. The detection and classification of price movements are based on objective data that describe the whole phenomenon. In general, technical and trading indicators are based on objective data and are usually reliable. Social media indicators are based on empirical measures obtained via deep learning algorithms trained with publicly available datasets: these datasets may carry intrinsic bias, which are in turn translated into classification errors of emotion and sentiment.

\section{Conclusions}
\label{sec:conclusions}

Several attempts have been made in the most recent literature to model and predict the erratic behaviour of prices or other market indicators of the major cryptocurrencies. Notwithstanding massive efforts devoted to this goal by many research groups, the analysis of cryptocurrency markets still remains one of the most debated and elusive tasks. Several aspects make grappling with this issue so complicated. For instance, due to its relatively young age, the cryptocurrency market is very dynamic and fast-paced. The emergence of new cryptocurrencies is a routine event, resulting in unexpected and frequent changes in the makeup of the market itself. Moreover, the high price volatility of cryptocurrencies and their ‘virtual’ nature are at the same time a blessing for investors and traders, and a curse for any serious theoretical and empirical modelling, with huge practical implications. The study of such a young market, whose price behaviour is still largely unexplored, has fundamental repercussions not only in the scientific arena but also for investors and main players and stakeholders in the crypto-market landscape. 

In this paper, we aimed to assess whether the addition of social and trading indicators to the ``classic” technical variables would lead to practical improvements in the classification of price changes of cryptocurrencies considering hourly and daily frequencies. This goal was achieved implementing and benchmarking a wide array of deep learning techniques, such as \emph{Multi-Layer Perceptron} (MLP), \emph{Multivariate Attention Long Short Term Memory Fully Convolutional Network} (MALSTM-FCN), \emph{Convolutional Neural Network} (CNN) and \emph{Long Short Term Memory} (LTMS) neural networks. We considered in our analysis the two main cryptocurrencies, Bitcoin and Ethereum, and we analysed two models: a restricted model, considering only technical indicators, and an unrestricted model that includes social and trading indicators. 

In the restricted analysis, the model that achieved the best performance, in terms of accuracy, precision, recall, and f1-score, is MALSTM-FCN with an average f1-score of $54\%$ for Bitcoin and the CNN for Ethereum with hourly frequency. For the unrestricted case the best result is achieved by the LSTM neural network for both Bitcoin and Ethereum with an average accuracy of $83\%$ and $84\%$ respectively. The most important finding for the hourly frequency classification for the unrestricted model is that the addition of trading and social indicators to the model leads to an effective improvement in the average accuracy, precision, recall, and f1-score. We have verified that this finding is not the result of a statistical fluctuation, since all the implemented models yielded the same achievements. For the same reason, we can exclude that the results depend on the particular implemented algorithm. Finally, for the daily classification, the best classification performance has been achieved by MALSTM-CNF for Ethereum with $99\%$ of accuracy when using the restricted model including only technical indicators. For Bitcoin, the best results are achieved by MLP with f1-score of $55\%$ and accuracy of $60\%$ with the unrestricted model including social media indicators and technical indicators, in this case, we consider f1-score and accuracy for Bitcoin because of the slightly unbalanced class distribution described in Section 3.4. For the daily frequency classification, we can see that in general technical indicators alone perform better in the classification of next day price movements. The more indicators we add to the model, the more the performance decreases. 

Another general result is that the accuracy, precision, recall, and f1-score for daily classification of Ethereum price movements are far better than those for Bitcoin. Our results show that with a specific design and fine-tuning of deep learning architecture, it is possible to achieve high performance in the classification of price changes of cryptocurrencies.

\bibliographystyle{splncs04}
\bibliography{paper}

\end{document}